\documentclass[prd,amsmath,amssymb,floatfix]{revtex4}

\usepackage{graphicx}

\begin{document}

\title{An Extended Model for the Evolution of Prebiotic Homochirality: A Bottom-Up Approach to the Origin of Life}
\author{Marcelo Gleiser}
\email{gleiser@dartmouth.edu}

\author{Sara Imari Walker}
\email{sara.i.walker@dartmouth.edu}

\affiliation{Department of Physics and Astronomy, Dartmouth College
Hanover, NH 03755, USA}

\begin{abstract}
A generalized autocatalytic model for chiral polymerization is investigated in detail. Apart from enantiomeric cross-inhibition, the model allows for the autogenic (non-catalytic) formation of left and right-handed monomers from a substrate with reaction rates $\varepsilon_L$ and $\varepsilon_R$, respectively. The spatiotemporal evolution of the net chiral asymmetry is studied for models with several values of the maximum polymer length, $N$. For $N=2$, we study the validity of the adiabatic approximation often cited in the literature. We show that the approximation obtains the correct equilibrium values of the net chirality, but fails to reproduce the short time behavior. We show also that the autogenic term in the full $N=2$ model behaves as a control parameter in a chiral symmetry-breaking phase transition leading to full homochirality from racemic initial conditions. We study the dynamics of the $N\rightarrow \infty$ model with symmetric ($\varepsilon_L=\varepsilon_R$) autogenic formation, showing that it only achieves homochirality for $\varepsilon < \varepsilon_c$, where $\varepsilon_c$ is an $N$-dependent critical value. For 
$\varepsilon \leq\varepsilon_c$ we investigate the behavior of models with several values of $N$, showing that the net chiral asymmetry grows as $\tanh(N)$. We show that for a given symmetric autogenic reaction rate, the net chirality and the concentrations of chirally pure polymers increase with the maximum polymer length in the model. We briefly discuss the consequences of our results for the development of homochirality in prebiotic Earth and possible experimental verification of our findings.
\end{abstract}


\maketitle

\section{\textbf{Introduction}}

Life is dependent upon biomolecular spatial asymmetry. Even though the origin of this asymmetry remains unknown \cite{Fitz}, it is frequently argued that life could neither exist nor originate without molecular chirality \cite{Bonner}. Under this view, the question of the origin of homochirality is intrinsic to that of the origin of life: if we want to understand how life first emerged on Earth (or possibly elsewhere), we must understand how homochirality emerges from basic chiral building blocks (enantiomers). 

The mystery here may not be so much {\it which} chirality but {\it why} chirality: although the sugars found in DNA and RNA are right-handed and protein amino acids are left-handed, laboratory syntheses yield racemic mixtures \cite{Blackmond04}. A pertinent exception is the reaction studied by Soai and coworkers, which demonstrates bifurcation to a single handedness \cite{Soai}. Although the Soai reaction has probably little relevance to life on early Earth,
it exhibits the features required of a successful model: autocatalysis and enantiomeric cross-inhibition fed by enzymatic enhancement.

Inspired by Frank's pioneering work \cite{Frank53}, Sandars recently proposed an autocatalytic model including enantiomeric cross-inhibition \cite{Sandars03}. Sandars' model has been further investigated in work by Brandenburg and collaborators
\cite{BM, BAHN}, Wattis and Coveney \cite{WC}, and Gleiser and Thorarinson \cite{GT, G}. One of the model's shortcomings is that enantiomeric feedback is generated by the longest polymer chain in the model. This requires somewhat unrealistic initial conditions, as they must include initial asymmetric concentrations of the maximum length polymers with no intermediate length polymers present in the reactor pool ({\em i.e.}, requiring the presence of $L_N$ and $R_N$ with some asymmetry and not requiring the presence of $L_n$ or $R_n$ for $n < N$). In a more realistic bottom-up approach, initial conditions should depend only on the substrate and possibly on a very small asymmetry between the two types of monomers, perhaps caused by random fluctuations in concentration \cite{Dunitz}. At an even more basic level, one might desire the system to start with no polymers, monomers, and even no substantial concentrations of substrate, so long as autogenic production of monomers is active. Such non-catalytic production terms for monomers of both chiralities were briefly discussed by Wattis and Coveney \cite{WC}, Saito and Hyuga \cite{SH}, and by Brandenburg {\em et al.} \cite{BM, BAHN} but have not yet been thoroughly explored.

In the present work, we extend Sandars' model by allowing for spontaneous monomer production from the substrate and for spatial dependence of the concentrations. We study in detail the spatiotemporal dynamics of the reaction network for several values of maximum polymer length $N$, including the $N\rightarrow \infty$ limit. For the $N=2$ (or reduced) model, we investigate the validity of the adiabatic approximations employed by Brandenburg and Multam\"aki \cite{BM}, Gleiser and Thorarinson \cite{GT}, and by Gleiser \cite{G}. We show that the introduction of autogenic monomer production influences the evolution of the net chirality by inducing a chiral symmetry-breaking phase transition. For models with $N>2$, we obtain the net chirality as a function of $N$, showing that it is sensitive to the autogenic reaction rates, $\varepsilon_{L(R)}$, for left (right)-handed monomers. For symmetric autogenic reactions,
($\varepsilon_{L(R)}=\varepsilon$), we show that above a critical value $\varepsilon_c$ no net chirality is produced: the racemizing autogenic term overwhelms the chirality-inducing enantiomeric cross-inhibition.

The paper is organized as follows. In section \ref{GM}, {\em The General Model}, we introduce the polymerization model and the relevant rate equations. The key features of autocatalysis, enantiomeric cross-inhibition, and enzymatic enhancement are discussed. In section \ref{reducedmodel}, {\em The Reduced Model}, we describe the reduced ($N=2$) model and the interpretation of the net chirality as an order parameter satisfying an effective potential. We also test the validity of the adiabatic approximations and describe the dynamics of a chiral symmetry-breaking transition. In section \ref{highN}, {\em The $N \rightarrow \infty$ Model}, we investigate the limit $N \rightarrow \infty$. For symmetric autogenic reactions, we obtain the phase diagram for the net chirality as a function of $\varepsilon$, showing that above a critical value $\varepsilon_c$ homochirality is unattainable.  We also compute the average polymer length as a function of $\varepsilon$, showing that as $\varepsilon\rightarrow \varepsilon_c$, large chiral chains are strongly suppressed. We then examine the spatiotemporal dynamics for several models with $ N \leq 10$ and $N \rightarrow \infty$ in section \ref{cp}, {\em Numerical Results: Maximum Polymer Length and Chiral Asymmetry}. We obtain an expression for the net chirality as a function of maximum polymer length $N$ as $\varepsilon$ varies. We also demonstrate that larger polymers achieve higher chiral purity. In section \ref{end}, {\em Conclusion}, we present a brief summary of our results.

\section{\textbf{The General Model}}\label{GM}

Sandars \cite{Sandars03} proposed a model for prebiotic homochirality whereby homochiral polymers evolve from a gradual build up of chiral monomers. The model includes the key features for chiral amplification,  enantiomeric cross-inhibition and enzymatic enhancement. In this section, we will present our generalization of Sandar's model, and discuss how it compares with other models proposed recently in the literature. Given that little is known of the specifics of prebiotic chemistry, it is important to examine the effects of plausible features expected to be present in realistic models.

\subsection{Modeling Polymerization}\label{MP}
Consider a chirally pure chain of $n$ left-handed monomers, $L_n$. It can grow on either end by addition of a left or right-handed monomer, $L_1$ or $R_1$, with rate coefficients $k_S$ and $k_I$, respectively. The attachment of the ``wrong'' monomer, in this case a right-handed monomer, inhibits that end of the polymer from further growth. This process is referred to as enantiomeric cross-inhibition: addition of a monomer of the opposite chirality to one end of a growing chain terminates further growth on that end of the chain \cite{Joyce}. The reaction network may be written as

\begin{align}\label{rxn network}
L_n + L_1 & \stackrel{2k_S}{\rightarrow} L_{n+1}, \nonumber \\
L_n + R_1 & \stackrel{2k_I}{\rightarrow} L_nR_1, \nonumber \\
L_1 + L_nR_1 & \stackrel{k_S}{\rightarrow} L_{n+1}R_1, \nonumber \\
R_1 + L_nR_1 & \stackrel{k_I}{\rightarrow} R_1L_nR_1, \nonumber \\
\end{align} 
supplemented by the complementary reactions obtained by exchanging $L \rightleftharpoons R$. 

In addition, the reaction network includes a substrate $S$ from which monomers of both chiralities are generated. The substrate may generate monomers spontaneously or through enzymatic enhancement by already existing polymers. The rate equations governing these processes are
\begin{eqnarray}\label{rxnNetwork1}
S \stackrel{\varepsilon_L}{\rightarrow} L_1, ~~~~~ 
S \stackrel{k_C p C_L}{\rightarrow} L_1,~~~~
S \stackrel{k_C q C_R}{\rightarrow} R_1, \nonumber \\ 
S \stackrel{\varepsilon_R}{\rightarrow} R_1, ~~~~
S \stackrel{k_C p C_R}{\rightarrow} R_1, ~~~~ 
S \stackrel{k_C q C_L}{\rightarrow} L_1,
\end{eqnarray}
where $\varepsilon_L$ and $\varepsilon_R$ are the rate coefficients for direct (autogenic or non-catalytic) production of monomers from the substrate, $k_C$ is the rate coefficient for enantiomeric feedback, $C_L$($C_R$) denote the enzymatic enhancement of left(right)-handed monomers, and  the coefficients $p$ and $q$ are given by, $p= \frac{1}{2}(1+f)$ and $q= \frac{1}{2}(1-f)$, where $f$ is the fidelity of the enzymatic reactions. Note that for complete fidelity ($f=1$), $q = 0$, and the two rightmost processes in eq. \ref{rxnNetwork1} do not occur. 

For simplicity, Sandars chose $C_L = L_N$ and $C_R = R_N$, where $N$ is the maximum polymer length. However, given that the exact functional dependence is not known, alternative suggestions have been made, such as $C_L = \sum nL_n$ and $C_R = \sum nR_n$ \cite{WC, BAHN} (henceforth WC and BAHN, respectively). Here, we allow for polymers of all lengths to participate equally in the autocatalytic feedback and choose $C_L = \sum L_n$ and $C_R = \sum R_n$. In contrast to the case where the enzymatic enhancement is dependent only on the longest $N$-length polymers, this choice allows  homochiral polymers of all lengths to participate in the autocatalytic feedback and removes any dependence on the specific cutoff value of $N$. This choice also allows the network to start with initial conditions which do not require the artificial presence of small, left-right asymmetric concentrations of the highest order polymer, a limitation of Sandar's model. In a bottom-up approach, long chains should be products of a gradual build-up: it is unlikely that they would be present {\it ab initio}. 

The rate coefficients $\varepsilon_L$ and $\varepsilon_R$ describe the production of monomers from substrate without the need for catalytic feedback. Given that little is known of the details of prebiotic chemistry, it is important to consider the effects of such terms on the reaction kinetics. Furthermore, chiral monomers could also have been fed to the early Earth environment from outer-space, as supported from findings of chiral amino acids in, {\it e.g.}, the Murchinson meteorite \cite{Cronin98,Engel97}.
Similar terms have previously been introduced by WC, Saito and Hyuga \cite{SH} (henceforth SH), and by BAHN, although WC and SH did not allow for biased production. In the model of WC, a single non-catalytic production term $\varepsilon = \varepsilon_L = \varepsilon_R$ is introduced. WC consider only $\varepsilon \ll 1$ to allow for simplified initial conditions ({\em i.e.}, starting with no substrate, monomers, or polymers present but with a continuous source term for the substrate) and did not study the effect of this term on the dynamics of the model or its equilibrium solutions, as we do here. BAHN introduced the non-catalytic production terms $C_{0L}$ and $C_{0R}$. Comparing to our model, one finds $\varepsilon_L = k_C C_{0L}$ and $\varepsilon_R = k_C C_{0R}$. BAHN found that the introduction of finite $C_{0L}$ or $C_{0R}$ with $C_{0L} \neq C_{0R}$ leads to imperfect but still efficient bifurcation in the feedback fidelity and to the absence of a racemic equilibrium solution. These authors suggest that homochirality is primarily the result of the instability of the racemic solutions which is barely modified by the presence of finite $C_{0L}$ or $C_{0R}$. As a consequence, most of BAHN's work assumed $C_{0L}=C_{0R}=0$. Here, we find that even in the case $\varepsilon_L = \varepsilon_R=\varepsilon$, the presence of these terms can, in fact, influence the instability of the racemic solutions: for $\varepsilon > \varepsilon_c(N)$, homochirality is impossible (as will be discussed, $\varepsilon_c(N)$ is a critical value which varies with maximum polymer length). A similar conclusion was reached by SH, who studied autocatalytic reactions in a simplified system with two chiral species. SH found that random production of chiral molecules from non-catalytic processes must occur to generate initial concentrations of chiral products and that there exists a critical value for the non-catalytic production rate: above the critical value, the system can obtain only fractional homochirality. We arrive at a qualitatively stronger conclusion: $\varepsilon_c(N)$ varies with the maximum length of polymers allowed in the model, $N$, and the net chirality for a given $\varepsilon < \varepsilon_c$ increases with increasing $N$.

Gleiser \cite{G} investigated the case where the bias originated in the enantiomeric enhancement terms through the definitions $Q_L = k_C (1+g/2)[S]C_L$ and $Q_R = k_C (1-g/2)[S]C_R$, where $g \ll 1$. Exploring the efficacy of both intrinsic and extrinsic influences through the parameter $g$, it was found that in the diffusive regime only a fairly large bias ($g\geq 10^{-6}$) could influence the evolution of reasonably-sized reactor pools toward homochirality in early Earth. In this regime, small sources of chiral symmetry breaking, such as parity violation in the weak interactions where $g\sim 10^{-17}$\cite{WNC0,KN85}, were ruled out.

With the above processes of deriving monomers (both non-catalytically and through enzymatic enhancement) present in the model, it is possible to start from racemic initial conditions and, through chiral amplification, arrive at a homochiral final state. One can imagine initially having zero concentrations of polymers, monomers, and substrate with the substrate being added continuously to the system starting at $t=0$. As pointed out by WC, if $\varepsilon_{L(R)} = 0$ no monomers are produced and polymerization is never initiated. (Models without autogenic production usually start with an unrealistic nonzero concentration of large polymers.) We therefore require the presence of non-zero autogenic production terms, $\varepsilon_L$ and $\varepsilon_R$, to produce the initial concentrations of monomers. In essence, they jump-start the reaction network.

There are two ways in which homochirality may be achieved once there is a nonzero concentration of substrate and possibly minute initial concentrations of monomers. If an explicit bias is introduced such that $\varepsilon_{L} \neq \varepsilon_{R}$, the system requires an initial concentration of only the substrate in order for chiral amplification toward the favored chirality to occur. If $\varepsilon_L = \varepsilon_R$, a small bias must be introduced between the initial concentrations of left and right-handed monomers. The needed asymmetry may be the result of random fluctuations in concentrations of large numbers of molecules \cite{Dunitz}, or it may be due to environmental disturbances \cite{GT, GTW}. We will explore both these cases, $\varepsilon_L = \varepsilon_R$ and $\varepsilon_{L} \neq \varepsilon_{R}$, in the context of the present paper.

\subsection{Polymerization Equations}
Taking the above into consideration, and accounting for the losses of $L_1$ and $R_1$ to lengthen growing chains, the set of reaction rate equations governing the various concentrations of a $N$-polymer network, are 
\begin{eqnarray}\label{N2eqs}
\frac{d[L_n]}{dt}&=& 2k_S [L_1][L_{n-1}] - 2 [L_n] (k_S[L_1] + k_I[R_1]), \\
\frac{d[R_n]}{dt}&=& 2k_S [R_1][R_{n-1}] - 2 [R_n] (k_S[R_1] + k_I[L_1]), \nonumber \\
\frac{d[L_n R_1]}{dt}&=& k_S [L_1][L_{n-1} R_1] + 2 k_I [R_1][L_n] - [L_n R_1] (k_S[L_1] + k_I[R_1]),\nonumber \\
\frac{d[R_n L_1]}{dt}&=& k_S [R_1][R_{n-1} L_1] + 2 k_I [L_1][R_n] - [R_n L_1] (k_S[R_1] + k_I[L_1]),\nonumber 
\end{eqnarray}\\
where $ 2<n \leq N $. We note that WC imposed the condition that $[L_nR_1]$ and $[R_nL_1]$ were inhibited from further growth by enantiomeric inhibition, so the last two eqs. above were not included in their model. Here, we assume that these chains can still grow on the opposite end from that attached to a ``wrong'' enantiomer. For $n=2$, we have the equations,
\begin{eqnarray}
\frac{d[L_2 R_1]}{dt}&=& 2 k_I [R_1][L_2] - [L_2 R_1] (k_S[L_1] + k_I[R_2]), \\
\frac{d[R_2 L_1]}{dt}&=& 2 k_I [L_1][R_2] - [R_2 L_1] (k_S[R_1] + k_I[L_2]). \nonumber
\end{eqnarray}

In the above equations, factors of $2$ arise when a monomer can attach to either end of a growing polymer. For dimers, one must discount this factor to account for the interaction of two single monomers. Thus, we must write
\begin{eqnarray}
\frac{d[L_2]}{dt}&=& k_S [L_1]^2 - 2 [L_2] (k_S[L_1] + k_I[R_1]), \\
\frac{d[R_2]}{dt}&=& k_S [R_1]^2 - 2 [R_2] (k_S[R_1] + k_I[L_1]). \nonumber
\end{eqnarray}
The evolution equations for left and right-handed monomers are,
\begin{eqnarray}
\frac{d[L_1]}{dt}&=& \varepsilon_L [S] + Q_L - \lambda_L [L_1], \\
\frac{d[R_1]}{dt}&=& \varepsilon_R [S] + Q_R - \lambda_R [R_1], \nonumber
\end{eqnarray}
where
\begin{eqnarray}
\lambda_L = 2k_S \sum_{n=1}^{N-1} [L_n] + 2k_I \sum_{n=1}^{\widetilde{N}} [R_n]+ k_S \sum_{n=2}^{N-1} [L_n R] + k_I
\sum_{n=2}^{\widetilde{N}} [R_n L], \nonumber \\
\lambda_R = 2k_S \sum_{n=1}^{N-1} [R_n] + 2k_I \sum_{n=1}^{\widetilde{N}} [L_n]+ k_S \sum_{n=2}^{N-1} [R_n L] + k_I
\sum_{n=2}^{\widetilde{N}} [L_n R]. \nonumber
\end{eqnarray}
$\lambda_L$ and $\lambda_R$ quantify the losses from monomer populations associated with attachment of a monomer to a growing polymer. $\varepsilon_L$ and $\varepsilon_R$ are rate coefficients for the direct production of monomers from the substrate. The choice of $\widetilde{N}$ is model dependent. In the model of Sandars, $\widetilde{N} = N-1$, and in the model of BAHN, $\widetilde{N} = N$. For the sake of simplicity, and to be consistent with other terms in the decay rates $\lambda_L$ and $\lambda_R$, we choose $\widetilde{N} = N-1$.

Assuming that the substrate is maintained by a source, $Q_S$, its evolution equation takes the form
\begin{eqnarray}
\frac{d[S]}{dt}&=& Q_S - (\varepsilon_L + \varepsilon_R)[S] -(Q_L + Q_R)
\end{eqnarray}
where $Q_L$ and $Q_R$ are the source terms for production of monomers from enzymatic enhancement and take the form 
\begin{eqnarray}
Q_L = k_C [S] (p C_L + q C_R), \\
Q_R = k_C [S] (p C_R + q C_L),	
\end{eqnarray}
where $p$ and $q$ are related to the fidelity as defined in section \ref{MP}. In this paper we set $f=1$ ({\em i.e.}, $q=0$ and $p=1$). As stated in section \ref{MP}, we choose $C_L = \sum_{n=2}^N [L_n]$ and $C_R = \sum_{n=2}^N [R_n]$. 

This completes the model. However, note that the rate equation for $[L_1R_1],$ which may be found in the polymerization model of Sandars, has been left out. Due to the effects of enantiomeric cross-inhibition, the further growth of this dimer is terminated; consequently, it does not affect the essential dynamics of the model and we have chosen not to include it since it has no bearing on the net chirality. Even so, it is important to acknowledge that hybrid dimers of this form are being created and thus are present in the system. One would need some method of removing these dimers, as well as longer chains containing the minority chirality from the reactor pool. Sandars suggested a recycling process involving a non-chiral intermediary to remove initial quantities of the unfavored enantiomer \cite{Sandars05}. We also note that even though most present-day biomolecules polymerize only unidirectionally \cite{NBAH}, given that we don't know the conditions of prebiotic chemistry, we chose to consider the more generic case where monomers can attach to both ends of a growing polymer (see eq.\ref {N2eqs}). The ensuing kinetics should be qualitative similar in the two cases, as the main differences will be that certain rate coefficients will be halved and that the equations for $[L_nR_1]$ and $[R_nL_1]$ are no longer necessary as these polymers do not grow.

In order to discuss the net chirality in later sections, we introduce the following variables of the enantiomeric excess
\begin{eqnarray}
\delta \equiv \frac{L_1 - R_1}{L_1 + R_1},~~ \theta \equiv \frac{\sum L_n - \sum R_n}{\sum L_n + \sum R_n}, ~~\eta_n \equiv \frac{L_n - R_n}{L_n + R_n}~(n\geq 2),
\end{eqnarray}
which describe different measures of chiral purity of homochiral polymers. In particular, $\theta$ accounts for the net chiral excess from all homochiral chains. A special case occurs in the truncated $N=2$ model with adiabatic approximations (discussed in the next section), where $\delta$ is the only measure we may 
use. The parameter $\eta_n$ describes the net chirality for chains of a specific length $n$. In the case $n = N$, this is equivalent to the measure of enantiomeric excess used by Sandars \cite{Sandars03}. In the following sections, we will compare the temporal and spatiotemporal evolution of systems for various values of $N$: $N =2,~5,~ 10$, and in the limit $N \rightarrow \infty$. We start by considering the simplest case, the $N=2$, or reduced, model.

\section{The Reduced Model} \label{reducedmodel}

We first consider the truncated $N=2$ model of Brandenburg and Multam\"aki \cite{BM} (henceforth BM), where the longest polymers formed are dimers. Apart from its simplicity, the attractiveness of this model relies on the fact that dimers play a similar catalytic role to that demonstrated in the Soai reaction, \cite {Blackmond04}. We will examine if such a truncation maintains the essential dynamics of a higher $N$ model. 

\subsection{The Reduced Model with Adiabatic Approximation}

BM simplified the reaction network for the $N=2$ system further by assuming that the rate of change of the concentrations of the substrate, $[S]$, and of dimers, $[L_2]$ and $[R_2]$, are much slower than that of the monomers, $[L_1]$ and $[R_1]$. This approximation is known as the adiabatic elimination of rapidly adjusting variables \cite{Haken}. We start by testing the validity of this approximation.

It proves convenient to introduce dimensionless symmetric and asymmetric variables, ${\cal S} \equiv X + Y$ and ${\cal A} \equiv X - Y$, respectively, where $X \equiv [L_1] (2k_S/Q_S)^{1/2}$ and $Y \equiv [R_1] (2k_S/Q_S)^{1/2}$. For $k_S/k_I =1$, after some algebra, the polymerization equations reduce to
\begin{eqnarray} \label{SA}
\lambda_0^{-1}\frac{d{\cal S}}{dt}&=&1-S^2,  \\
\lambda_0^{-1}\frac{d{\cal A}}{dt}&=&\frac{2{\cal S}(\gamma_L - \gamma_R+{\cal A})}{2(\gamma_L + \gamma_R){\cal
S}+{\cal S}^2+ {\cal A}^2} - {\cal S}{\cal A}, \nonumber
\end{eqnarray}
where the parameter $\lambda_0 \equiv (2k_SQ_S)^{1/2}$ has dimensions of inverse time, and we have introduced the parameters
\begin{eqnarray}\label{gammaeqn}
\gamma_L\equiv 2(\frac{2k_S}{Q_S})^{\frac{1}{2}}\frac{\varepsilon_L}{k_C}, ~~~  \gamma_R \equiv
2(\frac{2k_S}{Q_S})^{\frac{1}{2}}\frac{\varepsilon_R}{k_C}~.
\end{eqnarray}\\
For $\gamma_R = \gamma_L = 0$, the above equations reduce to the models of BM and of Gleiser and Thorarinson \cite{GT}.

The fixed point of the first equation in eqs. \ref{SA} is easily found to be ${\cal S} = 1$: the system tends quickly toward this value at time-scales of order $\lambda_0^{-1}$. With ${\cal S} = 1$, the equation for the chiral asymmetry becomes
\begin{eqnarray}
\lambda_0^{-1}\frac{d{\cal A}}{dt}&=&\frac{{\cal A} - {\cal A}^3 + 2 \gamma_L(1-{\cal A}) - 2 \gamma_R (1+{\cal
A})}{2(\gamma_R + \gamma_L)+ {\cal A}^2 +1}.
\end{eqnarray}
Writing the above in the more suggestive form, $\lambda_0^{-1} \dot{\cal A} = - \partial V / \partial {\cal A}$ (the dot denotes time derivative), one obtains the potential 
\begin{eqnarray}\label{potential}
V({\cal A})= \frac{{\cal A}^2}{2} - \ln[1 + {\cal A}^2 +2(\gamma_L + \gamma_R)] ~~~~~~~~~~~~~~~~~~~\nonumber \\
- \frac{2(\gamma_L- \gamma_R)}{\sqrt{1+2(\gamma_L + \gamma_R)}} \arctan\left[ \frac{{\cal A}}{\sqrt{1+2(\gamma_L + \gamma_R)}}\right]\nonumber \\.
\end{eqnarray}

\begin{figure}
\centering
\begin{minipage}[b]{0.5\textwidth}
\resizebox{85mm}{!}{\includegraphics[width=4in,height=3in]{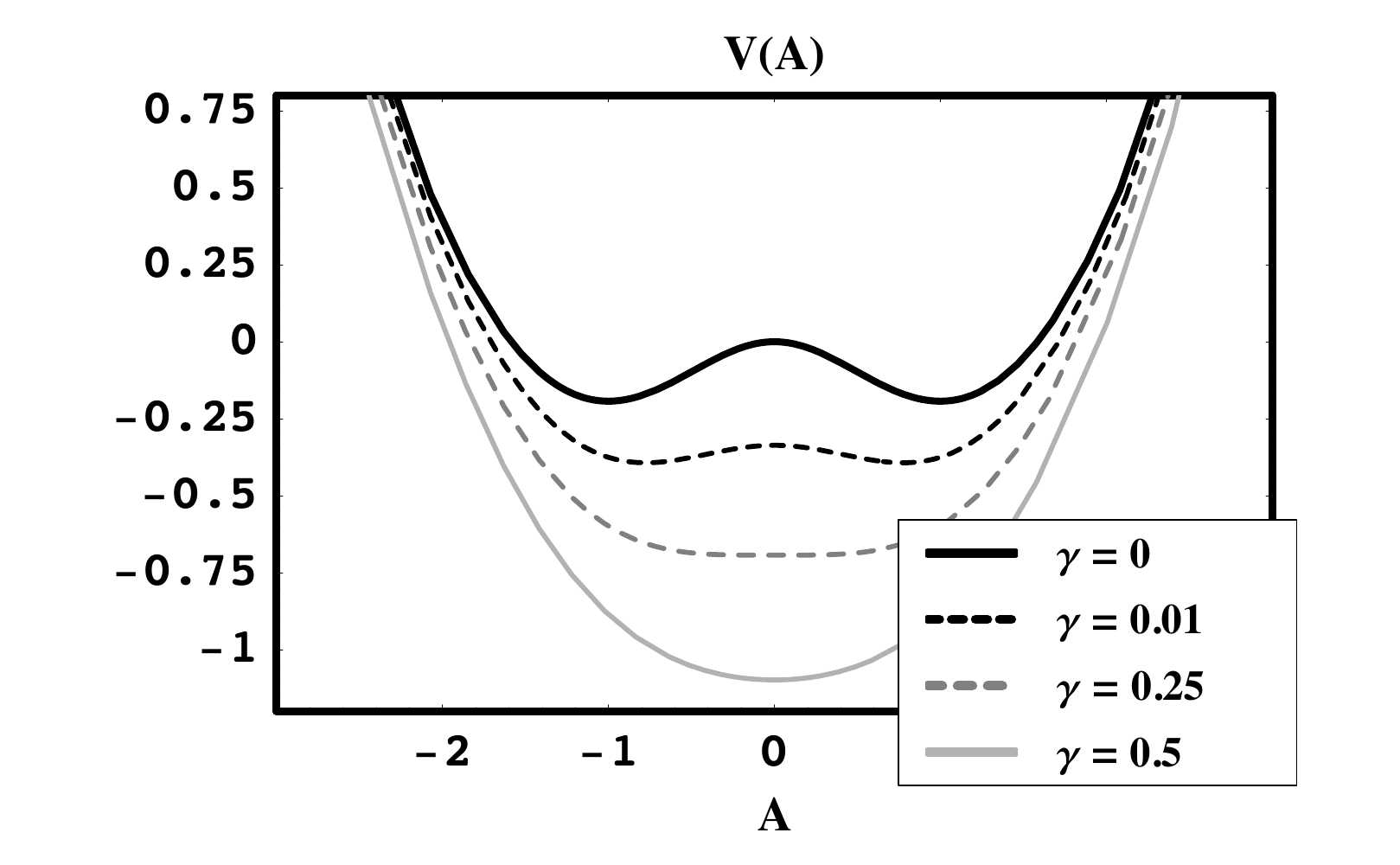}}
\end{minipage}%
\begin{minipage}[b]{0.5\textwidth}
\resizebox{85mm}{!}{\includegraphics[width=4in,height=3in]{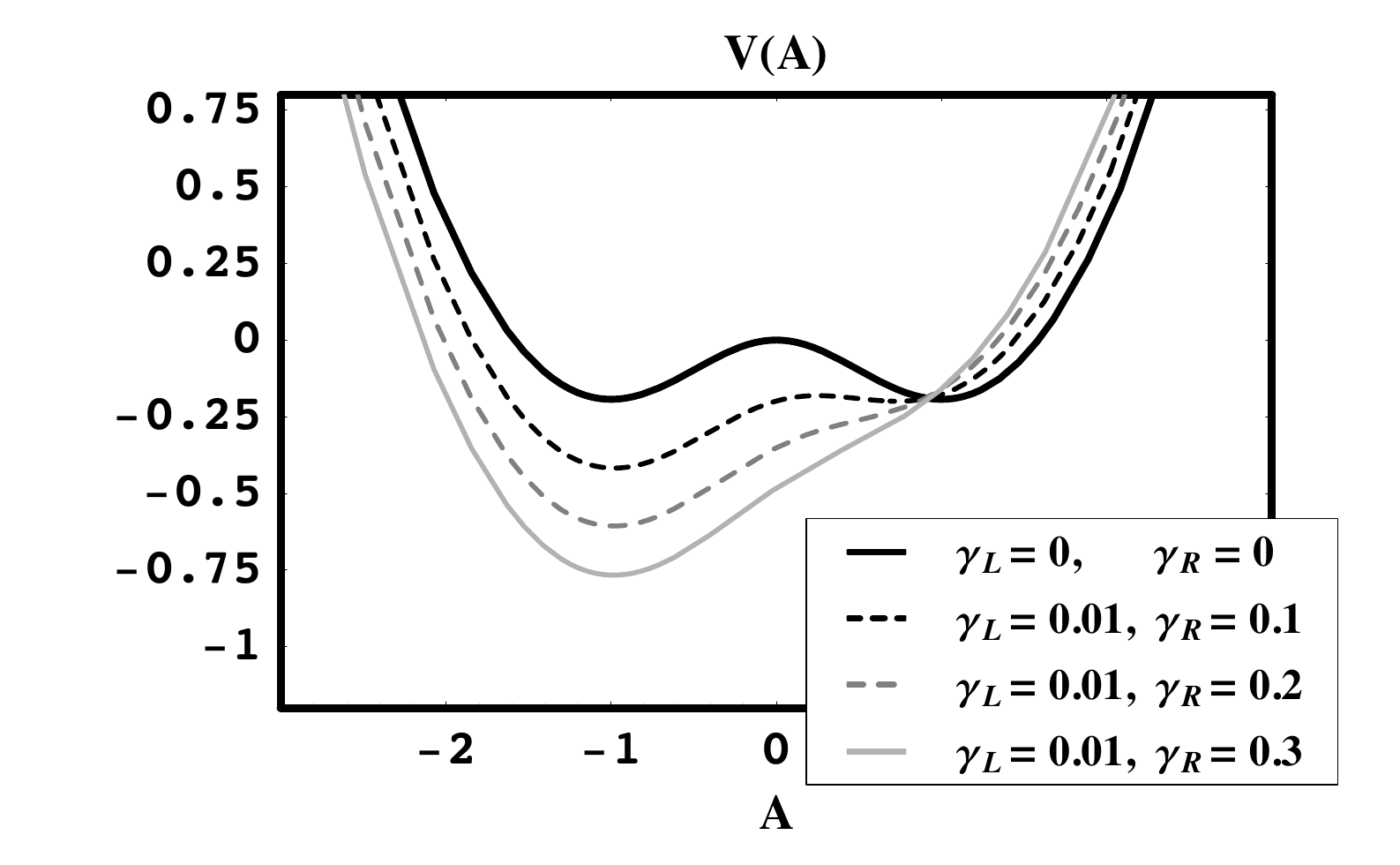}}
\end{minipage}%
\caption{Left: Potential $V({\cal A})$ for $\gamma_R = \gamma_L = \gamma$ for several values of $\gamma$.
Right: Potential for $\gamma_R \neq \gamma_L$ with varying asymmetry between the two parameters. In both graphs, the thick line corresponds to $ \gamma_R = \gamma_L =\gamma= 0$.} \label{fig:potential}
\end{figure}

The potential $V({\cal A})$ is symmetric under ${\cal A}\rightarrow -{\cal A}$ for $\gamma_L = \gamma_R=\gamma$, and asymmetric for $\gamma_L \neq \gamma_R$. Figure \ref{fig:potential} shows $V({\cal A})$ for the symmetric and asymmetric cases. 

Considering the symmetric case, $\gamma = \gamma_L = \gamma_R $, the potential has fixed points at ${\cal A} = 0$ and ${\cal A} = \pm \sqrt{1-4\gamma}$. The point ${\cal A}=0$ describes the racemic solution and is unstable when the chiral fixed points at ${\cal A} = \pm \sqrt{1-4\gamma}$ are real. The enantiomeric excess is given by
\begin{eqnarray}
\delta = \frac{{\cal A}}{{\cal S}} = \pm \sqrt{1-4\gamma}.
\label{gammacrit}
\end{eqnarray}
Clearly, an enantiomeric excess is possible only for $ \gamma < 0.25 $, and complete homochirality ($\delta = \pm 1$) occurs only for $\gamma = 0$. For $ \gamma > \gamma_c=0.25 $ a racemate (${\cal A}=0$) is the only stable solution. We see that $\gamma$ plays a similar quantitative role as the temperature in systems in the Ising universality class \cite{Goldenfeld}. We will further explore this analogy below.

As discussed previously, BAHN found that in cases with an asymmetry between non-catalytic production terms, there is no longer a perfectly racemic stable solution. For $\gamma_L$($\gamma_R$) large enough, the left(right)-handed chirality invariably will achieve dominance ({\it cf.} figure \ref{fig:potential} right). For the $N=2$ truncated model, the asymmetry between $\gamma_L$ and $\gamma_R$ plays a similar qualitative role to the biasing parameter $g$ introduced by Gleiser \cite{G}.

\subsection{Chiral Selection as a Phase Transition}

The form of the potential in eq. \ref{potential} suggests the possibility of chiral symmetry breaking initiated by a temporal or spatiotemporal varying parameter $\gamma(t)$ or $\gamma(\textbf{x},t)$, respectively. We treat spatial evolution in the following sections. Here, we investigate the parameter $\gamma$ and its qualitatively similar role to that played by temperature in phase transitions (see \cite{Goldenfeld}). We consider the symmetric case $\gamma = \gamma_L = \gamma_R$ and rewrite eqs. \ref{gammaeqn} as
\begin{eqnarray}\label{gammas}
\gamma \equiv 2(\frac{2k_S}{Q_S})^{\frac{1}{2}}\frac{\varepsilon}{k_C},
\end{eqnarray}
where $\varepsilon = \varepsilon_L = \varepsilon_R$.  

Consider a temporally varying parameter $\gamma(t)$. One can imagine this to occur with a change in temperature, pressure, or concentration of reactants as the system evolves. Starting with a large $\gamma$, for instance $\gamma(t=0)=1$, the production rate of monomers of both chiralities is relatively high ({\it i.e.}, $\varepsilon > k_C$, holding $k_S$ fixed). As $\gamma$ decreases, the rate of direct production of  monomers from the substrate slows, and enantiomeric feedback, governed by the reaction rate $k_C$, becomes the dominant mechanism for formation of monomers ({\it i.e.}, when $\varepsilon < k_C$). As studied extensively in previous work \cite{Sandars03, BAHN, WC}, enantiomeric feedback is the driving force that leads to eventual chiral bifurcation. Thus {\it the critical point in this model occurs when the dominant mechanism for monomer formation switches from autogenic production from the substrate to enantiomeric feedback}. The transition occurs at $\gamma = \gamma_c$ ($\gamma_c = 0.25$ for the symmetric potential in figure \ref{fig:potential}) and complete chiral separation occurs as $\gamma \rightarrow 0$. 

In this scenario, as $\gamma(t)$ passes through $\gamma_c$, the system undergoes a phase transition from a racemic state to a chiral state. One can think in analogy with a ferromagnet: as it is cooled through the Curie point, it settles into one of the two degenerate ground states (magnetization directions). The choice is determined by fluctuations about equilibrium. In the absence of an external magnetic field (the biasing factor, here equivalent to choosing $\gamma_L\neq \gamma_R$), the choice is random; there is an equal probability that either ground state will be chosen. The net chirality plays the role of the net magnetization and we observe domains of both chiralities emerge as $\gamma$ is lowered through $\gamma_c$.

If one then considers the parameter $\gamma$ to be both spatially and temporally dependent, $\gamma = \gamma(\vec{x},t)$ -- a realistic possibility given that, in general, the temperature, pressure, or concentrations are likely to be inhomogeneous in early planetary environments -- the phase transition can occur at different places at different times. For example, if $\gamma$ is dependent on temperature, then $\gamma(T)$ will be nonuniform, as the temperature itself may be a function of space and time ($T(\textbf{x},t)$). In this case, the phase transition may not occur everywhere at once and chirality will not be uniform throughout space. Chiral domains will then compete for dominance as discussed in Gleiser \cite{G}. The formation of chiral domains for a model phase transition will be illustrated below in the context of the full $N=2$ model.

\subsection{The Reduced Model Without Adiabatic Approximation}\label{N2noAdiabatic}

To lay out the groundwork for larger $N$, we explicitly write out the equations for the full $N=2$ model without adiabatic approximations. This will also allow us to test their validity. We define the dimensionless symmetric and asymmetric variables,
\begin{eqnarray}
{\cal S}_1 =(\frac{2k_S}{Q_S})^{\frac{1}{2}}([L_1] + [R_1]), ~~~
{\cal A}_1 =(\frac{2k_S}{Q_S})^{\frac{1}{2}}([L_1] - [R_1]),\nonumber \\
{\cal S}_2 =(\frac{2k_S}{Q_S})^{\frac{1}{2}}([L_2] + [R_2]), ~~~
{\cal A}_2 =(\frac{2k_S}{Q_S})^{\frac{1}{2}}([L_2] - [R_2]), \nonumber 
\end{eqnarray}
describing the evolution of the concentrations of monomers and dimers, and the dimensionless variable ${\cal \psi}=(\frac{2k_S}{Q_S})^{\frac{1}{2}}[S]$ describing the evolution of the substrate. The network equations are then (setting $k_I/k_S =1$),
\begin{eqnarray} \label{N2fullmodel}
\lambda_0^{-1}\frac{d \psi}{dt}&=&1-\frac{1}{2}\kappa(\gamma_L + \gamma_R)\psi - \kappa \psi {\cal S}_2,   \nonumber \\
\lambda_0^{-1}\frac{d{\cal S}_1}{dt}&=& \kappa \left(\frac{1}{2}(\gamma_L + \gamma_R )+ {\cal S}_2 \right)\psi - {\cal S}_1^2,   \nonumber \\
\lambda_0^{-1}\frac{d{\cal A}_1}{dt}&=& \kappa \left(\frac{1}{2}(\gamma_L - \gamma_R ) + {\cal A}_2 \right)\psi - {\cal S}_1 {\cal A}_1,   \nonumber \\
\lambda_0^{-1}\frac{d{\cal S}_2}{dt}&=& \frac{1}{4}({\cal S}_1^2 + {\cal A}_1^2) - {\cal S}_2 {\cal S}_1,\nonumber \\
\lambda_0^{-1}\frac{d{\cal A}_2}{dt}&=& \frac{1}{2}{\cal S}_1 {\cal A}_1 - {\cal A}_2 {\cal S}_1,
\end{eqnarray}
where, as before, $\lambda_0 = (2k_S Q_S)^{1/2}$. The parameters $\gamma_L$ and $\gamma_R$ were defined in eq. \ref{gammaeqn}, and we introduced the new parameter $\kappa \equiv \frac{k_C}{2 k_S}$. 

For the remainder of this work we assume $k_I = k_S = k_C$ ($\kappa = 0.5$): the rates of attachment of a monomer of the same or opposite chirality are equal, and these rates are in turn equal to the rate of catalytic production of monomers from the substrate by enantiomeric feedback. The other relevant rates that appear in our reaction network, $\varepsilon_L$ and $\varepsilon_R$, which describe non-catalytic monomer production, should be much slower \cite{WC, BAHN}. For example, WC chose $\varepsilon \approx 10^{-5}{\rm s}^{-1}$. To show that the condition $\varepsilon \ll 1$ is satisfied, we use the nominal values of $k_S \approx 10^{-25}$cm$^3$s$^{-1}$ and $Q_S \approx 10^{15}$cm$^{-3}$s$^{-1}$ with $k_C =k_S$ to find
\begin{eqnarray}
\gamma_{L(R)} = 2 \sqrt{2}\times 10^5 \varepsilon_{L(R)} ~ {\rm s}.
\end{eqnarray}
Noting that $\gamma$ should be dimensionless, we obtain $\varepsilon_{L(R)} \lesssim 10^{-6} {\rm s}^{-1}$ for $0 \leq \gamma < \frac{1}{4}$.  

\subsection{Comparison of Reduced Models: Testing the Adiabatic Approximation}

\begin{figure}
\centerline{\includegraphics[width=4in,height=3in]{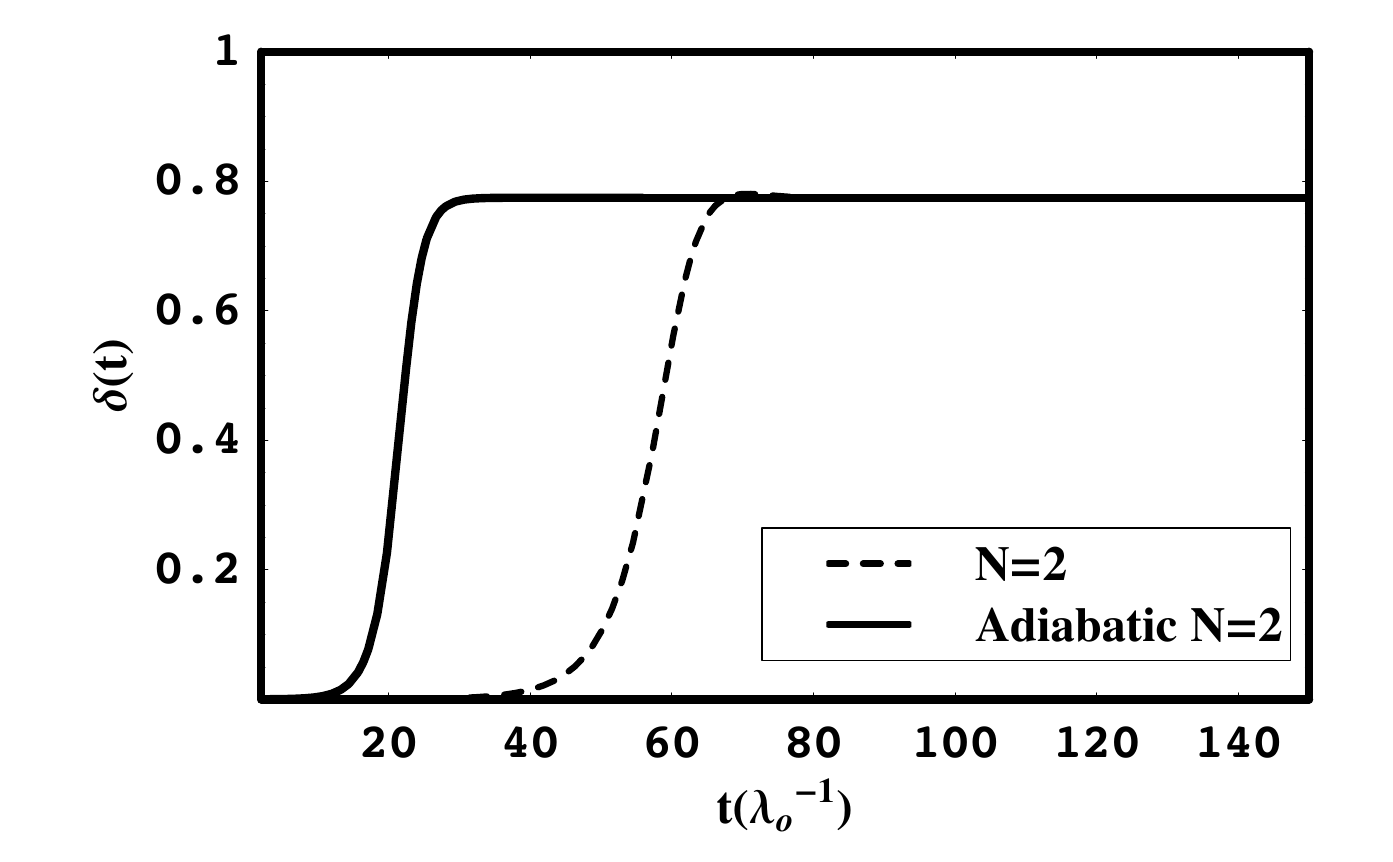}}
\caption{Time evolution of the net chirality for $N=2$ models with (continuous line) and without (dashed line) adiabatic approximations. Note that the asymptotic values of $\delta$ are equal, and given by eq. \ref{gammacrit}. } \label{fig:N2compare}
\end{figure}

In the previous subsections we introduced the $N=2$ model with and without adiabatic approximations. Since the adiabatic approximations have been used in several models \cite{BM, G, GT}, it is important to test their validity. In figure \ref{fig:N2compare}, we show the temporal evolution of $\delta$ for both models with $\gamma = 0.1$. The initial conditions were the same in both models: an asymmetry
$([L_1]-[R_1])|_{t=0}=10^{-5}$ in monomer concentrations. All ordinary differential equations in this work were solved numerically using Mathematica.  

If, as suggested by BM, adiabatic elimination does not affect the accuracy of steady-state solutions (where $d\delta/dt\rightarrow 0$ as $t\rightarrow \infty$), we should expect the asymptotic values to be equivalent. In figure \ref{fig:N2compare}, we see that the asymptotic solutions of both models are, indeed, equal ($\delta_{ss}=0.775$): not surprisingly, the difference between the two models is in the equilibration time-scales, which are considerably faster within the adiabatic approximation. We can therefore conclude that the adiabatic approximation is valid on long dynamical timescales. One can also verify analytically that $\gamma_c$ is the same for both models: both the adiabatic and non-adiabatic models will undergo a phase transition for the same values of the relevant parameters, $k_C$, $\varepsilon$, and $k_S$. (The phase diagram for the full $N=2$ model is shown in figure \ref{fig:ThetaNinf}). 

\subsection{Introducing Spatial Dependence: Chiral Phase Transition}\label{SD}

Spatial inhomogeneities in the concentrations of the many reactants will have a strong impact on the evolution of chirality. As we remarked earlier, the analogy with ferromagnets suggests that spatially extended domains of opposite chirality will form and compete for dominance \cite{BM,G}.
In what follows, we will introduce spatial dependence to the full $N=2$ model, and study its time evolution as the production rate of monomers from the substrate decreases in time.

The advantage of imposing the adiabatic approximation to the $N=2$ model is most obvious once spatial dependence is introduced: the system is elegantly reduced to an effective scalar field theory where the field determines the net chirality in a given volume \cite{BM}, playing the same role as the magnetization in an Ising ferromagnet. However, as we demonstrated in the previous subsection (see figure \ref{fig:N2compare}), the adiabatic approximation changes the effective equilibration time-scale, thus affecting the overall evolution of the reaction network. We thus move on to introduce spatial dependence to the full $N=2$ model, following the usual procedure in the phenomelogical treatment of phase transition dynamics \cite{Langerrev, Gunton}, by writing the total time derivatives in eqs. \ref{N2fullmodel} as $d/dt \rightarrow \partial / \partial t - k \nabla^2$, where $k$ is the diffusion coefficient. Typical values for $k$ are $k= 10^{-9}$m$^2$s$^{-1}$ for molecular diffusion in water and $k= 10^{-5}$m$^2$s$^{-1}$ for air.

The equations are made dimensionless by introducing the time and space variables, $t_0 \equiv \lambda_0 t$ and $x_0 \equiv x\sqrt{\lambda_0/k}$. Dimensionful values are then recovered for a particular choice of the parameters $k_S$, $Q_S$, and $k$. Using the nominal values of $k_S \approx 10^{-25}$cm$^3$s$^{-1}$, $Q_S \approx 10^{15}$cm$^{-3}$s$^{-1}$, and $k$ for water, one obtains $\lambda_0 \cong \sqrt{2} \times 10^{-5}$s$^{-1}$, which yields to $t \cong 2.3 \times 10^{-3}t_0$y and $x \cong 8.5 \times 10^{-3}x_0$m. 

To investigate the dynamical evolution of the net chirality as it undergoes a chiral phase transition, we  simulated the evolution of the net chirality in a reactor pool by allowing $\gamma$ to decrease linearly from $\gamma_0 = 0.3 > \gamma_c$ to $\gamma = 0$, writing
$\gamma(t) = \gamma_0(1-t/t_f)$, where $t_f$ is chosen so that $\gamma(t_f)=0$. For the full $N=2$ model, the net chirality is given by,
\begin{align}
\theta = \frac{{\cal A}_1 + {\cal A}_2}{{\cal S}_1 + {\cal S}_2},
\end{align}
where ${\cal A}_1$, ${\cal A}_2$, ${\cal S}_1$, and ${\cal S}_2$ were defined in section \ref{N2noAdiabatic}.

To generate small initial spatial fluctuations in the concentrations, we coupled the system to an external environment via a stochastic spatiotemporal Langevin equation. Rewriting eqs. \ref{N2fullmodel} we have, 
\begin{eqnarray} \label{N2space}
\lambda_0^{-1}\left(\frac{\partial \psi~}{\partial t} - k \nabla^2 \psi~ \right)&=& 1-\frac{1}{2}\kappa(\gamma_L + \gamma_R)\psi - \kappa \psi {\cal S}_2 + \xi(\textbf{x}, t),   \nonumber \\
\lambda_0^{-1}\left(\frac{\partial{\cal S}_1}{\partial t} - k \nabla^2 {\cal S}_1 \right)&=& \kappa \left(\frac{1}{2}(\gamma_L + \gamma_R )+ {\cal S}_2 \right)\psi - {\cal S}_1^2  + \xi(\textbf{x}, t),   \nonumber \\
\lambda_0^{-1}\left(\frac{\partial{\cal A}_1}{\partial t} - k \nabla^2 {\cal A}_1 \right)&=& \kappa \left(\frac{1}{2}(\gamma_L - \gamma_R ) + {\cal A}_2 \right)\psi - {\cal S}_1 {\cal A}_1  +  \xi(\textbf{x}, t),   \nonumber \\
\lambda_0^{-1}\left(\frac{\partial{\cal S}_2}{\partial t} - k \nabla^2 {\cal S}_2\right)&=& \frac{1}{4}({\cal S}_1^2 + {\cal A}_1^2) - {\cal S}_2 {\cal S}_1  +  \xi(\textbf{x}, t),\nonumber \\
\lambda_0^{-1}\left(\frac{\partial{\cal A}_2}{\partial t} - k \nabla^2 {\cal A}_2 \right)&=& \frac{1}{2}{\cal S}_2 {\cal A}_2 - {\cal A}_2 {\cal S}_1  + \xi(\textbf{x}, t) ,
\end{eqnarray}
where $\xi({\textbf x}, t)$ is a stochastic force with zero mean ($\left\langle \xi \right\rangle =0$) and two-point correlation function $\left\langle \xi ({\textbf x'}, t') \xi ({\textbf x}, t)\right\rangle = a^2 \delta({\textbf x'} - {\textbf x})\delta(t'-t)$, and $a^2$ measures the strength of the external influence in units of (length)$^d$time, where $d$ is the number of spatial dimensions.

\begin{figure}
\centerline{\includegraphics[width=4in,height=3in]{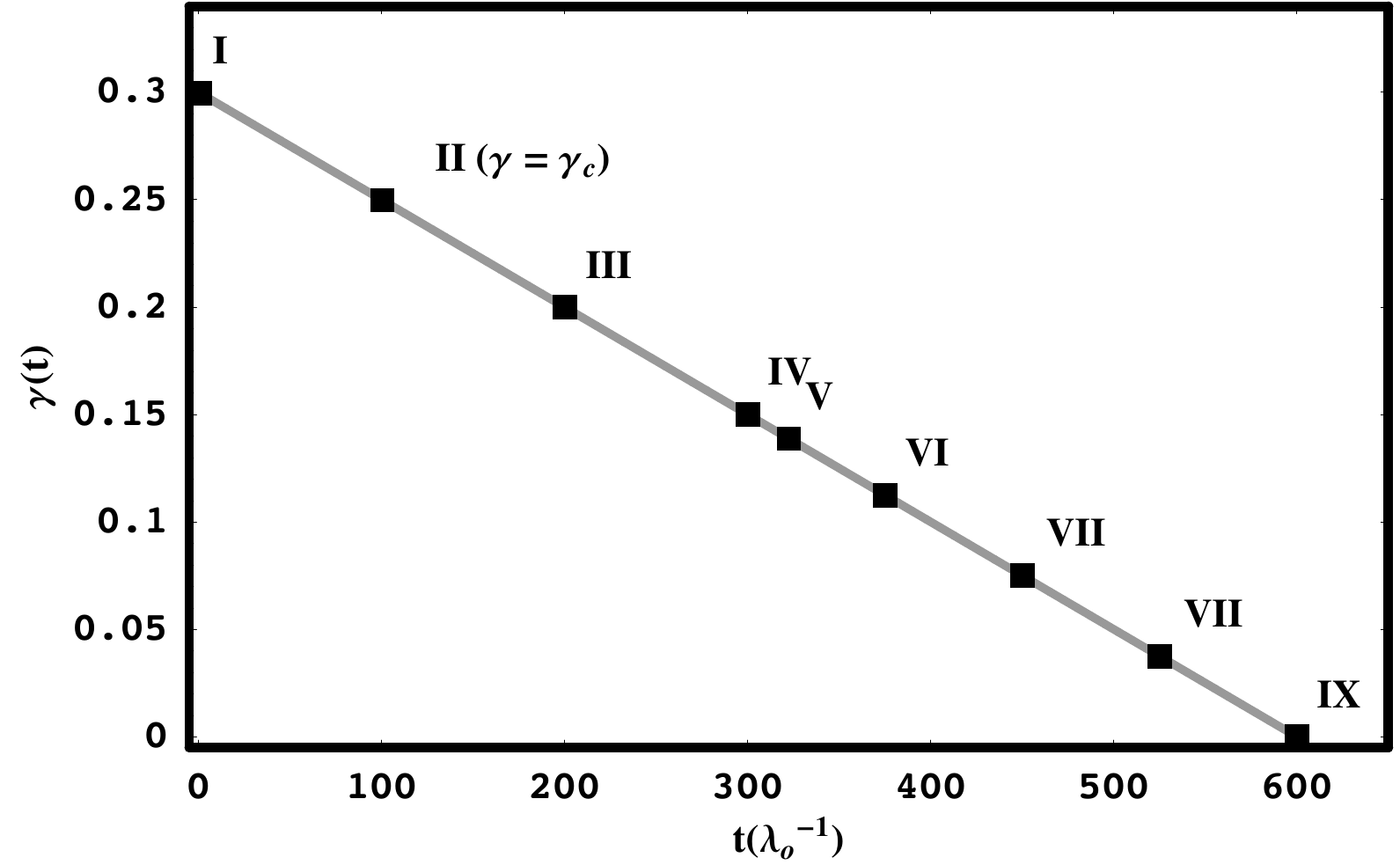}}
\caption{Evolution of $ \gamma(t)=0.3(1-t/600)$ for the full $N=2$ model. The Roman numerals in the figure correspond to the 2d snapshots shown in figure \ref{fig:2DChiral}.} \label{fig:gammat}
\end{figure}

The equations dictating the evolution of the reaction network (eqs. \ref{N2space}) were solved with a finite-difference method in a $1024^2$ grid with $\delta t = 0.005$ and $\delta x = 0.2$, and periodic boundary conditions. In $2$d, this corresponds to simulating a shallow pool of linear dimension $l \approx 200$cm. The reactor pool was initially set with a near-racemic distribution of monomers and dimers (as would occur in the case that the system started from a small amount of subtrate and was permitted to evolve for a period of time) with $\left\langle \theta(t=0) \right\rangle \sim 10^{-4}$. (The angled brackets denote spatial averaging, $\langle\dots\rangle = (1/V)\int\dots dV$.) The functional dependence for $\gamma(t)$ is shown in figure \ref{fig:gammat}. In addition to $\gamma$ linearly decreasing with time, we also imposed a linear time dependence on the environmental influence such that $a^2(t=0)=a_0^2=0.02$ and $a(t_f=600)=0$. In other words, $\gamma(a) = \gamma_0(a/a_0)$, thus establishing a relationship between the substrate ability to produce monomers and the external environment. One may think of $a(t)$ as an effective temperature: as $a(t)\rightarrow 0$, the autogenic production of monomers from the substrate ceases. Note that we chose $a_0$ to be sufficiently small such that the environmental influences do not affect the dynamical evolution of the net chirality. As demonstrated by Gleiser and Thorarinson \cite{GT}, for $a$ above a critical value $a_c$, chiral symmetry is restored. In the $N=2$ model with adiabatic approximations, $a_c^2=1.15$ in 2d, and $a_c^2=0.65$ in 3d. A recent study of the critical influence of the environment on homochirality can be found in ref. \cite{GTW}.

\begin{figure}
\centerline{\includegraphics[width=4in,height=3in]{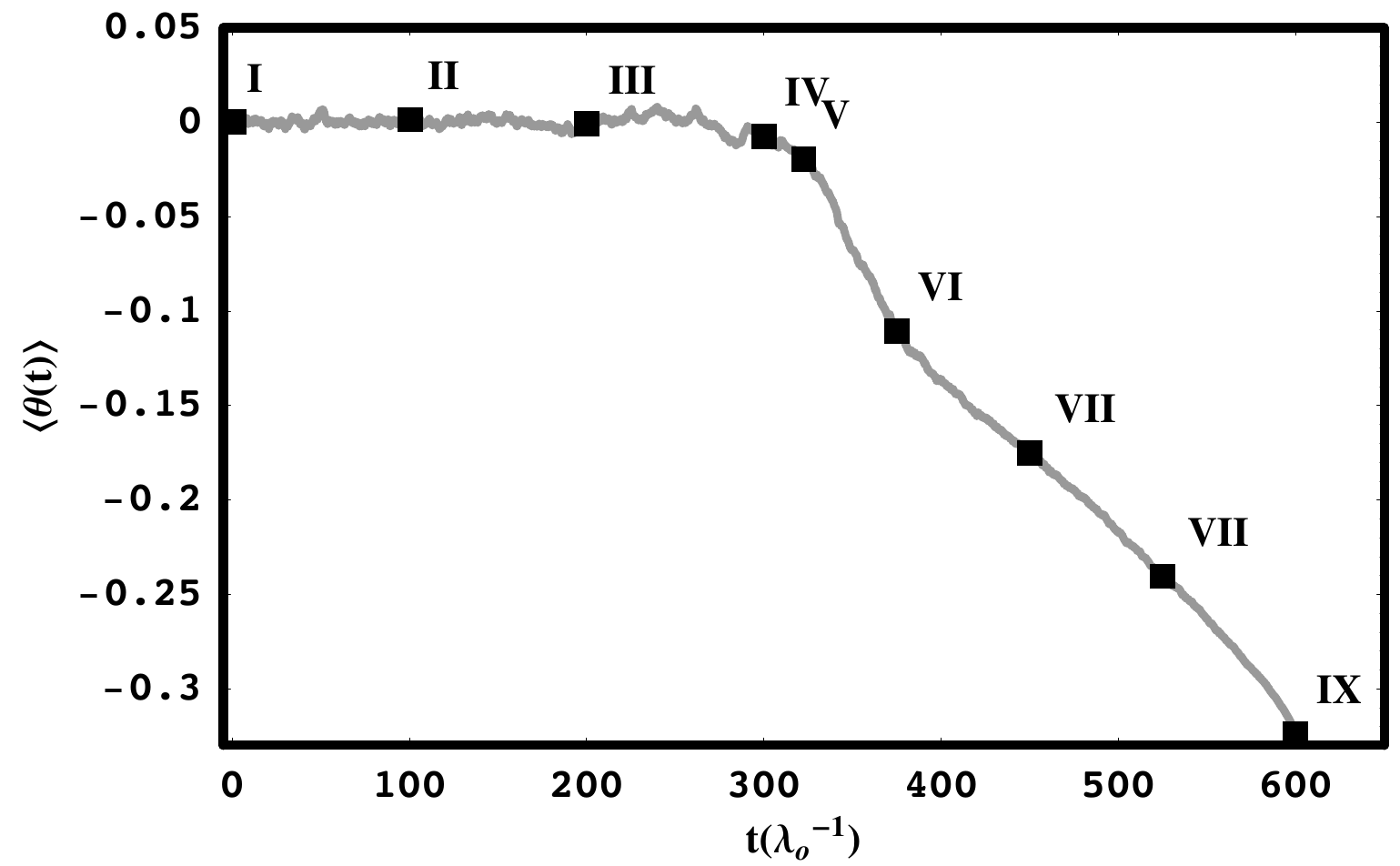}}
\caption{Evolution of the spatially-averaged net chirality $ \left\langle \theta(t) \right\rangle$ for the full$~N=2$ model with $\gamma = \gamma(t)$ as specified in figure \ref{fig:gammat}. The Roman numerals in the figure correspond to the snapshots shown in figure \ref{fig:2DChiral}.} \label{fig:2DChiralThetaAve}
\end{figure}

\begin{figure}
\centerline{\includegraphics[width=6in,height=7.125in]{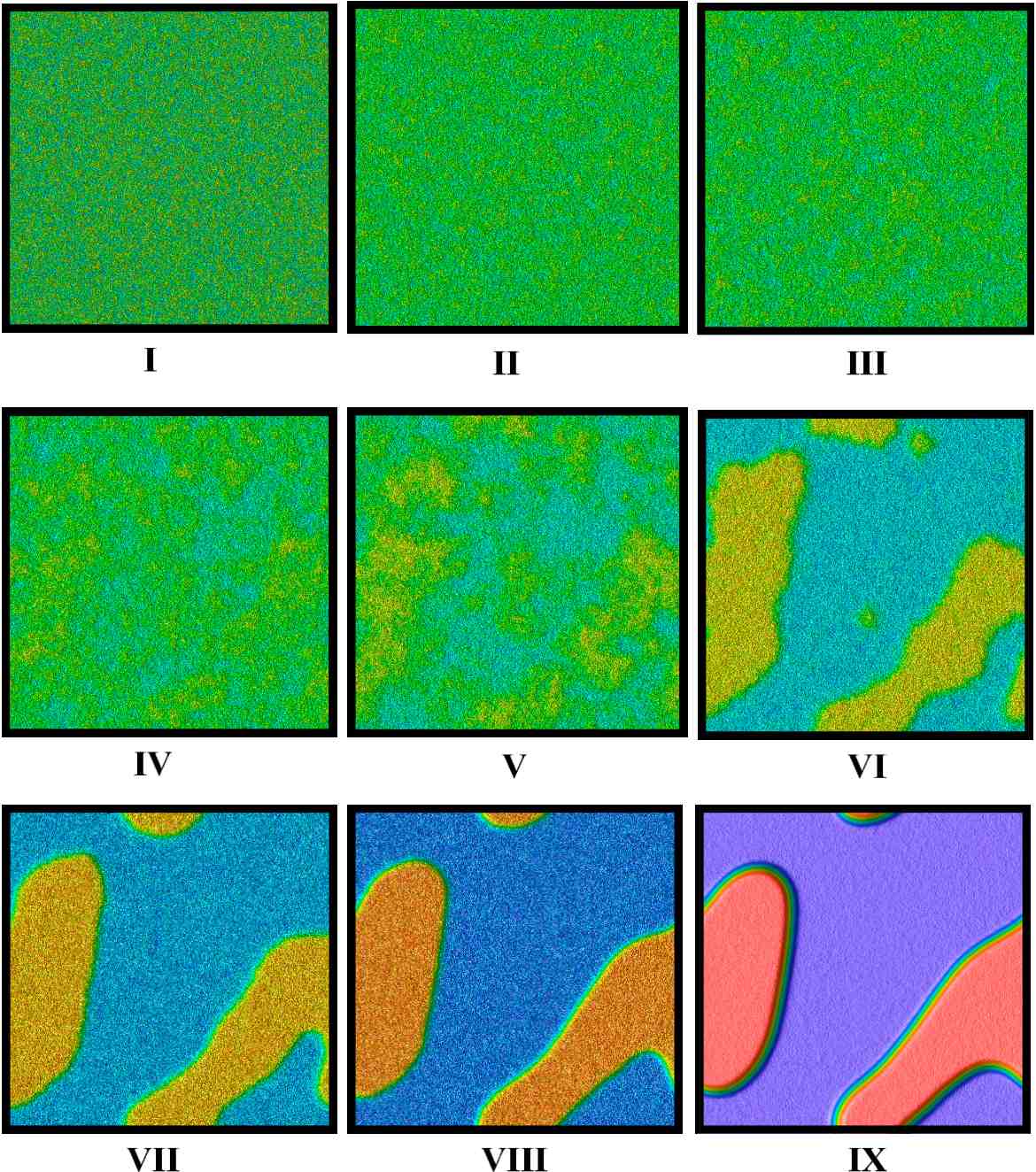}}
\caption{Evolution of net chirality for the full $N=2$ model, with $\gamma = \gamma(a)$, where $a$ is a measure of environmental influences. Each snapshot depicts the net chirality $\theta(x,y)$ at a time specified in figure \ref{fig:2DChiralThetaAve}. Note that $\gamma_c = 0.25$ is reached in {\rmfamily \bfseries II}: for $\gamma < \gamma_c$ domains of opposite chirality begin to form and compete for dominance.} \label{fig:2DChiral}
\end{figure}

The temporal evolution of the spatially-averaged net chirality $\left\langle  \theta(t) \right\rangle$ is shown in figure \ref{fig:2DChiralThetaAve} and two-dimensional snapshots of the reactor pool are shown in figure \ref{fig:2DChiral}. In looking at the evolution of $\left\langle \theta(t) \right\rangle$, we see that there is a long period of time where the net chirality is nearly racemic, even after $\gamma$ drops below $\gamma_c$. However, once $\gamma$ becomes as small as $0.15$, larger values of opposing chiralities become possible and we see that the net chirality evolves toward the chirally pure phase $\left\langle \theta(t) \right\rangle = -1$.

In snapshot {\textbf I} of figure \ref{fig:2DChiral}, $\gamma = 0.3$ and $\left\langle \theta(t) \right\rangle \approx 10^{-5}$. With $\gamma > \gamma_c$ the tendency is for the system to remain a near-uniform racemate. 
In snapshots {\textbf II} and {\textbf III} , $\gamma = \gamma_c =0.25$ and $\gamma = 0.2$, respectively. Even though $\gamma < \gamma_c$, the phase space available to $\theta$ is still very small, so we do not yet observe any chiral domains forming. In snapshot {\textbf IV}, $\gamma = 0.15$ and we begin to observe chiral domains as $\theta(t,x,y)$ is permitted to take on larger values. As $\gamma$ is decreased still further, in snapshots {\textbf V} through {\textbf IX}, the domains coarsen and the magnitudes of both the red ($L$) and blue ($R$) phases increase. Once domains of both chiralities are present, the curvature pressure of the walls will begin to affect the dynamics \cite{G}. Here the  red ($L$) phase does not percolate the lattice so it will continue to shrink in size; eventually, only the blue ($R$) phase will remain and homochirality will be achieved, completing the phase transition.

\section{The $N \rightarrow \infty$ Model}\label{highN}

To study the full set of polymerization equations as they might apply to a realistic biochemistry, one must consider the reaction network in the limit of large $N$. Fortunately, as pointed out by WC, the set of infinite rate equations may be reduced exactly to a closed system of only a handful of equations, significantly reducing the complexity of the problem. 

To investigate the infinite set of rate equations for our model, we introduce the new variables 
\begin{align}
[L] \equiv \sum_{n=2}^\infty [L_n],~~~~~~~ [R] \equiv \sum_{n=2}^\infty [R_n], \nonumber \\
[U] \equiv \sum_{n=2}^\infty [L_n R_1],~~~ [T] \equiv \sum_{n=2}^\infty [R_n L_1], \nonumber \\
[G] \equiv \sum_{n=2}^\infty n[L_n],~~~~ [D] \equiv \sum_{n=2}^\infty n[R_n].
\end{align}
The reaction network may be written succinctly in terms of these new variables. Setting $k_I/k_S =1$, the entire network reduces to (for $N \rightarrow \infty$)
\begin{eqnarray}
\frac{d{[S]}}{dt}&=&Q_S-(\varepsilon_L + \varepsilon_R)[S] - k_C [S]([L] + [R]),  \nonumber \\
\frac{d{[L_1]}}{dt}&=& (\varepsilon_L + k_C[L])[S] - 2k_S[L_1]\left([L_1]+[R_1]+[R]+[L]+\frac{1}{2}([U]+[T])\right),\nonumber \\
\frac{d{[R_1]}}{dt}&=& (\varepsilon_R + k_C[R])[S] - 2k_S[R_1]\left([L_1]+[R_1]+[R]+[L]+\frac{1}{2}([U]+[T])\right),\nonumber \\
\frac{d{[L]}}{dt}&=&k_S[L_1]^2 - 2k_S[R_1][L], \nonumber \\
\frac{d{[R]}}{dt}&=&k_S[R_1]^2 - 2k_S[L_1][R], \nonumber \\
\frac{d{[U]}}{dt}&=&2k_S[R_1][L] - k_S[R_1][U], \nonumber \\
\frac{d{[T]}}{dt}&=&2k_S[L_1][R] - k_S[L_1][T], \nonumber \\
\frac{d{[G]}}{dt}&=&2k_S[L_1]^2 + 2k_S[L_1][L] - 2k_S[R_1][G], \nonumber \\
\frac{d{[D]}}{dt}&=&2k_S[R_1]^2 + 2k_S[R_1][R] - 2k_S[L_1][D].
\end{eqnarray}
Similar to the procedure for the reduced model, we introduce symmetric and asymmetric dimensionless variables
\begin{eqnarray}\label{Ninfdefs} 
{\cal S} =(\frac{2k_S}{Q_S})^{\frac{1}{2}}([L_1] + [R_1]), ~~~
{\cal A} =(\frac{2k_S}{Q_S})^{\frac{1}{2}}([L_1] - [R_1]),\nonumber \\
{\cal N} =(\frac{2k_S}{Q_S})^{\frac{1}{2}}([L] + [R]), ~~~~~
{\cal \eta} =(\frac{2k_S}{Q_S})^{\frac{1}{2}}([L] - [R]), \nonumber \\
{\cal M} =(\frac{2k_S}{Q_S})^{\frac{1}{2}}([U] + [T]), ~~~~~
{\cal \mu} =(\frac{2k_S}{Q_S})^{\frac{1}{2}}([U] - [T]), \nonumber \\
{\cal P} =(\frac{2k_S}{Q_S})^{\frac{1}{2}}([G] + [D]), ~~~~~
{\cal \rho} =(\frac{2k_S}{Q_S})^{\frac{1}{2}}([G] - [D]),
\end{eqnarray}
along with the dimensionless variable ${\cal \psi}=(\frac{2k_S}{Q_S})^{\frac{1}{2}}[S]$ describing the substrate. The network in terms of these new variables becomes
\begin{eqnarray}\label{Ninfeqs}
\lambda_0^{-1}\frac{d{\cal \psi}}{dt}&=&1-\frac{1}{2}\kappa(\gamma_L + \gamma_R)\psi - \kappa \psi {\cal N},  \nonumber \\
\lambda_0^{-1}\frac{d{\cal S}}{dt}&=&\frac{1}{2}\kappa(\gamma_L + \gamma_R)\psi +\kappa \psi {\cal N}
-{\cal S}({\cal S} + {\cal N} + \frac{1}{2} {\cal M}),  \nonumber \\
\lambda_0^{-1}\frac{d{\cal A}}{dt}&=&\frac{1}{2}\kappa(\gamma_L - \gamma_R)\psi +\kappa \psi {\cal \eta}-{\cal
A}({\cal S} + {\cal N} + \frac{1}{2} {\cal M}), \nonumber \\
\lambda_0^{-1}\frac{d{\cal N}}{dt}&=& \frac{1}{4}({\cal S}^2 + {\cal A}^2) - \frac{1}{2}({\cal S}{\cal N} -
{\cal A}{\cal \eta}), \nonumber \\
\lambda_0^{-1}\frac{d{\cal \eta}}{dt}&=&\frac{1}{2}{\cal S}{\cal A} - \frac{1}{2}({\cal S}{\cal \eta} -
{\cal A}{\cal N}), \nonumber \\
\lambda_0^{-1}\frac{d{\cal M}}{dt}&=& \frac{1}{2}({\cal S}{\cal N} - {\cal A}{\cal \eta}) - \frac{1}{4}({\cal
S}{\cal M}- {\cal A}{\cal \mu}),  \nonumber \\
\lambda_0^{-1}\frac{d{\cal \mu}}{dt}&=& \frac{1}{2}({\cal S}{\cal \eta} - {\cal A}{\cal N}) - \frac{1}{4}({\cal
S}{\cal \mu}- {\cal A}{\cal M}) \nonumber \\
\lambda_0^{-1}\frac{d{\cal P}}{dt}&=& \frac{1}{2}({\cal S}^2 + {\cal A}^2) + \frac{1}{2}({\cal S}{\cal N}+{\cal A}\eta) - \frac{1}{2}({\cal S}{\cal P}- {\cal A}\rho),  \nonumber \\
\lambda_0^{-1}\frac{d{\cal \rho}}{dt}&=& {\cal S}{\cal A} + \frac{1}{2}({\cal S}\eta + {\cal A}{\cal N})-\frac{1}{2}({\cal S}\rho - {\cal A}{\cal P}).
\end{eqnarray}\\
The parameters $\gamma_L$, $\gamma_R$, and $\kappa$ are as defined in section \ref{N2noAdiabatic}.

\begin{figure}
\centerline{\includegraphics[width=4in,height=3in]{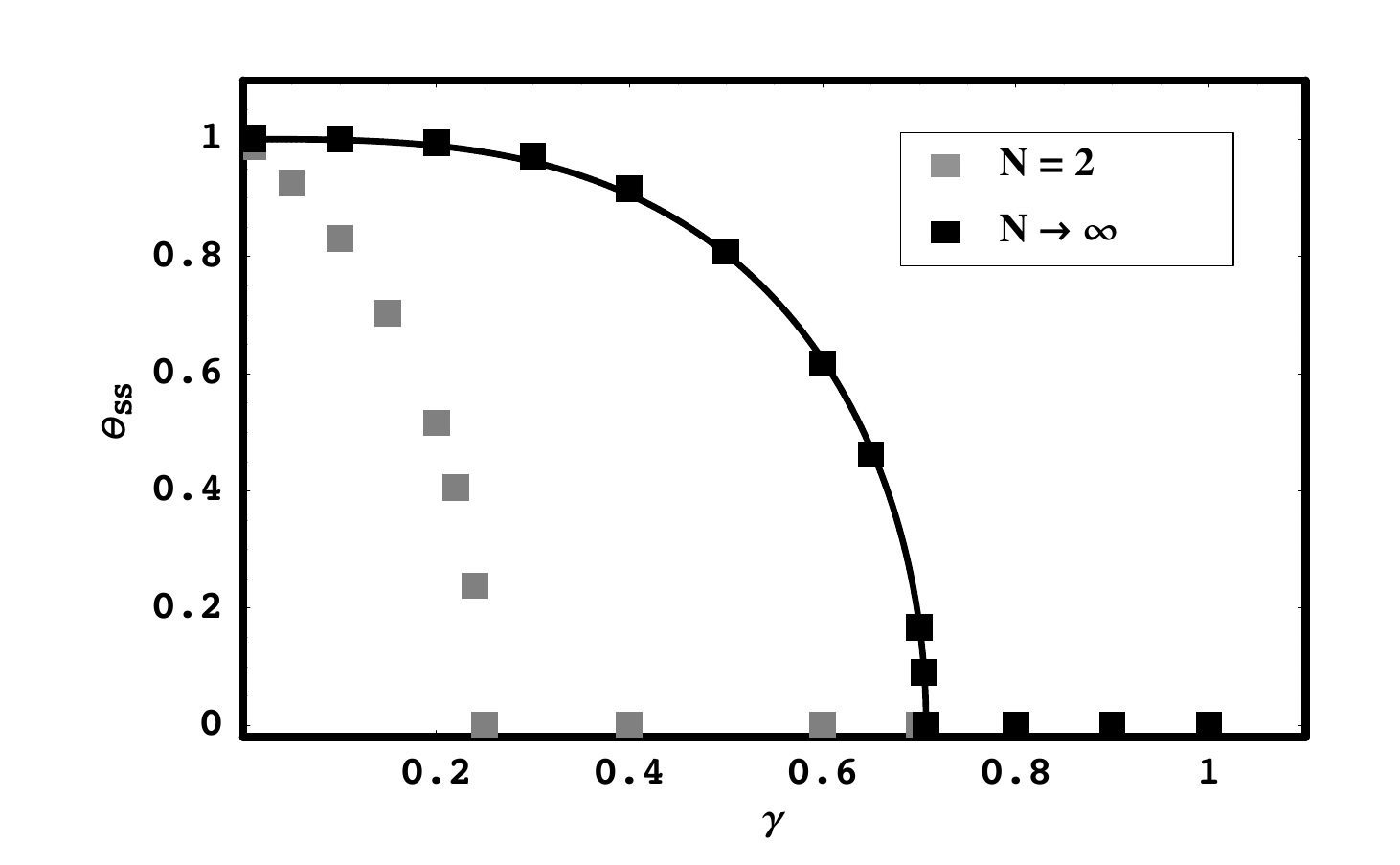}}
\caption{Phase diagram for the steady state chirality $\theta_{ss}$ as a function of the parameter $\gamma$ in the limit $N \rightarrow \infty$ (black squares), where $\gamma = \gamma_L=\gamma_R$. Note that $\gamma_c \mid _{N \rightarrow \infty} = \sqrt{2}/2$. Also included is the phase diagram for $N=2$ (grey squares), with $\gamma = \gamma_L=\gamma_R$, for comparison. The continuous line is the fit $\theta_{ss} = (1 - (\gamma_c/\gamma)^3)^{1/2}$} \label{fig:ThetaNinf}
\end{figure}

Note that in terms of these new variables the net chirality is simply
\begin{eqnarray}
\theta = \frac{{\cal A} + {\cal \eta}}{{\cal S} + {\cal N}}.
\label{thetanet}
\end{eqnarray}
As shown in figure \ref{fig:ThetaNinf}, it is possible to construct a phase diagram for the steady state value of the net chiral asymmetry, $\theta_{ss}$, by obtaining the equilibrium solutions to eqs. \ref{Ninfeqs} for varying $\gamma$. We note that in the $N\rightarrow \infty$ limit the only possible steady-state solution for $\gamma=0$ is racemic (not shown in figure \ref{fig:ThetaNinf}): there are, however, runaway solutions that, in spite of never reaching a steady state, still reach homochirality. This can be seen from eq. \ref{thetanet}: even though $\eta$ and ${\cal N}$ keep growing (the runaway behavior), their ratio approaches unity as ${\cal A}$ and ${\cal S}$ approach small values. As WC noted, in this case the average polymer length goes to infinity, not a very desirable feature. This supports including autogenic reactions in more realistic models. But, as we shall see, these reactions cannot be too efficient or they will lead invariably to a racemate, also not a desirable feature.

In figure \ref{fig:ThetaNinf}, we show the phase diagram for the $N=2$ and the $N \rightarrow \infty$ models. We have shown above that $\gamma_c|_{N=2}=0.25$. The critical value $\gamma_c \mid _{N \rightarrow \infty} = \frac{\sqrt{2}}{2}$ was found empirically: it is the only value of $\gamma$ that yields a set of solutions which are all racemic and contains no imaginary solutions. (For $\gamma < \gamma_c$ chiral solutions exist and, for $\gamma > \gamma_c$, there exist imaginary steady-state solutions. Only at $\gamma =\gamma_c$ we should see no chiral and no imaginary steady-state solutions). 

\begin{figure}
\centerline{\includegraphics[width=4in,height=3in]{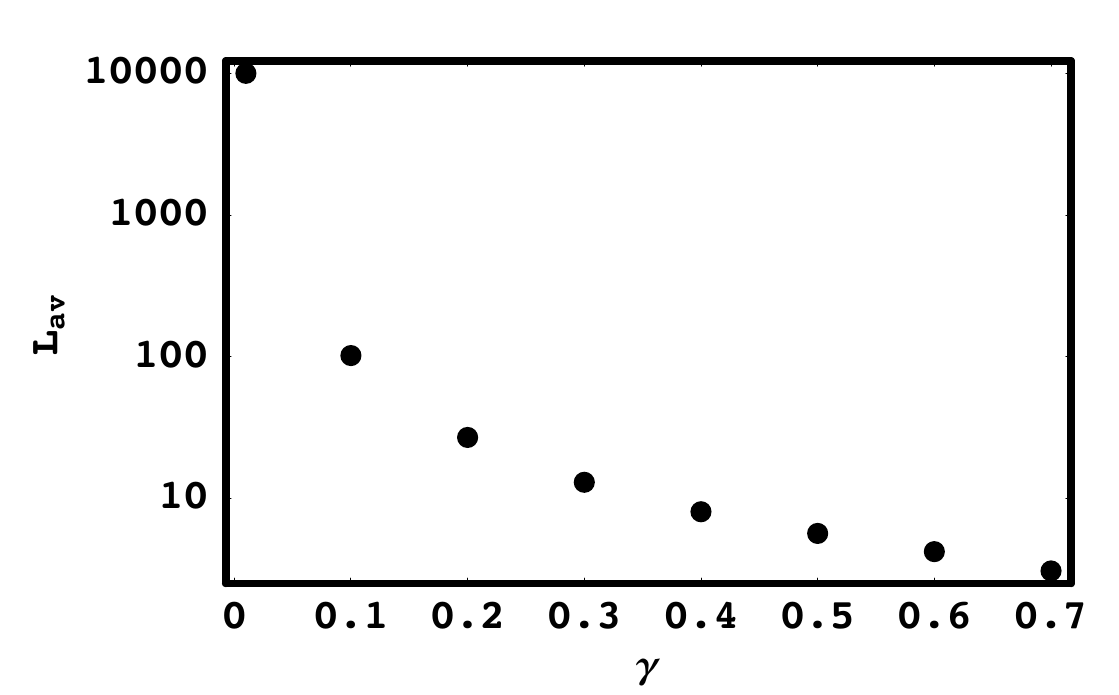}}
\caption{Average polymer lengths as a function of $\gamma$ for $N \rightarrow \infty$. SS indicates that the values are for the steady state solutions.} \label{fig:PolymerLengths}
\end{figure}

The average polymer length is defined as,
\begin{eqnarray}
L_{av} \equiv \frac{\sum n[L_n] + \sum n[R_n]}{\sum [L_n] + \sum [R_n]}=\frac{{\cal P}}{{\cal N}}, 
\end{eqnarray}
where ${\cal P}$ and ${\cal N}$ where introduced in eqs. \ref{Ninfdefs}. 
As mentioned previously, for $\gamma = 0$ the average polymer length diverges. However, for $\gamma$ non-zero, $L_{av}$ takes on finite values. The steady-state average polymer lengths for $0.01 \leq \gamma \leq 0.7$ are shown in figure \ref{fig:PolymerLengths}. For racemic solutions, $L_{av} = 3$. This is consistent with the results of figure \ref{fig:PolymerLengths}: as $\gamma$ approaches $\gamma_c$, the system approaches a racemate and the average polymer is a trimer. 

\section{Spatiotemporal Dynamics of Polymerization: Maximum Polymer Length and Chiral Asymmetry}\label{cp}

In the previous sections, we have alluded to a relationship between the maximum polymer length in the reactor pool, $N$, and the steady-state value of the chiral asymmetry, $\theta_{ss}$. In particular, we ask the question: how does the presence of a monomer-producing substrate with rate $\gamma$ affect the reactor pool's ability to achieve homochirality for different models with maximum polymer length $N$? In other words, for a given $\gamma$, is it easier or harder to achieve homochirality with increasing polymer length?  

The phase diagrams for $N=2$ and $N \rightarrow \infty$ of figure \ref{fig:ThetaNinf} show that, in principle, systems with large $N$ achieve homochirality for a wider range of parameter space. Inspecting figure \ref{fig:ThetaNinf}, it can be seen that for $N \rightarrow \infty$ significant homochirality, $\theta \geq 0.9$, is achievable for relatively large values of $\gamma$, $\gamma \lesssim 0.4$, as compared to $\gamma \lesssim 0.06$ for $N=2$.  

\begin{figure}
\centerline{\includegraphics[width=4in,height=3in]{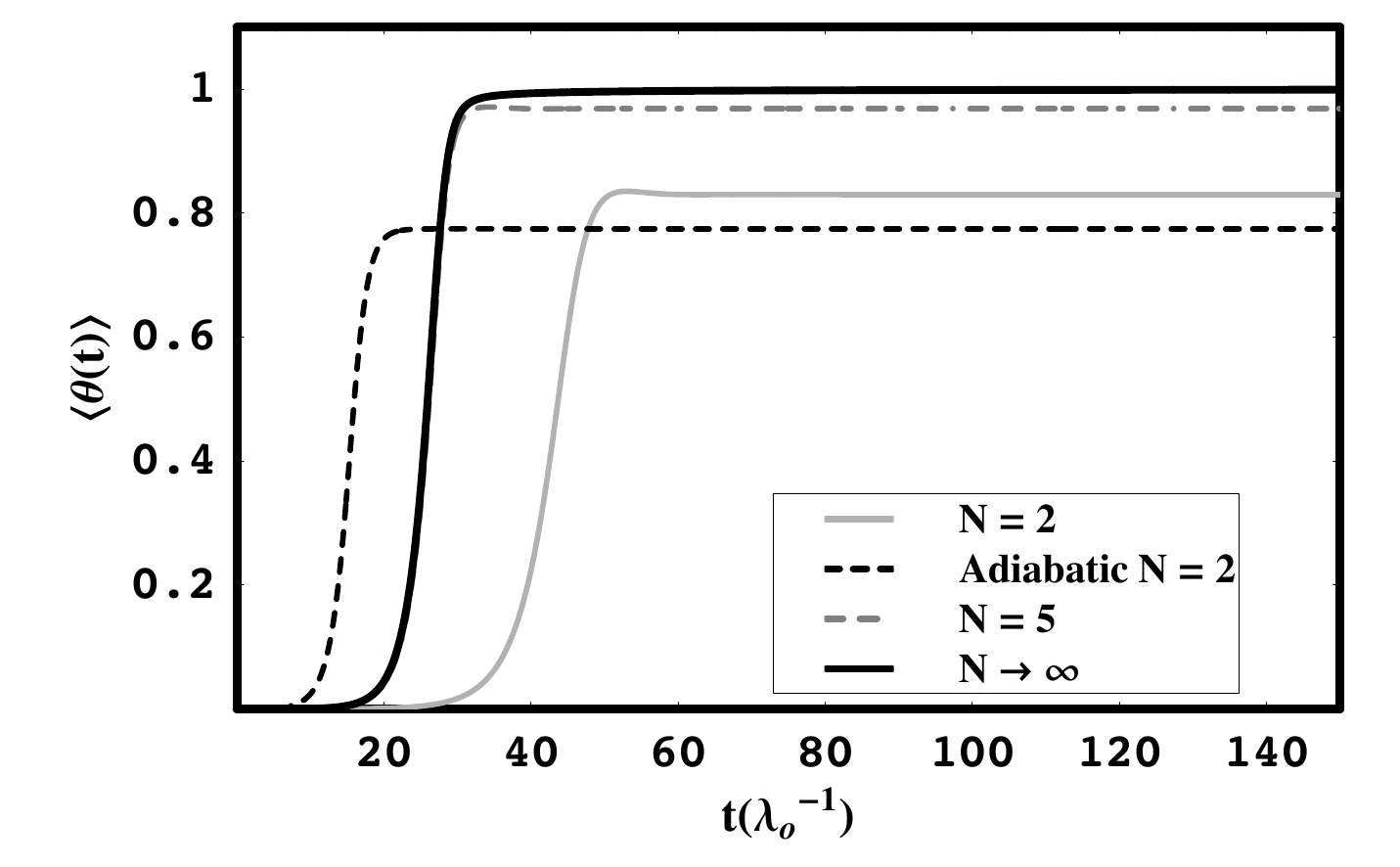}}
\caption{Temporal evolution of spatially-averaged net chirality for $N=2$ (grey line), $N=2$ with adiabatic approximation (dashed black line), $N=5$ (dashed grey line), and $N \rightarrow \infty $ (continuous black line) for $\kappa =0.5$ and $\gamma = 0.1$. (Note that for $N=2$ with adiabatic approximations, $\left\langle \delta(t)\right\rangle$ is plotted instead of $\left\langle \theta(t)\right\rangle$.)} \label{fig:Ncompare}
\end{figure}

To investigate the spatiotemporal dynamics of the reaction network for different $N$, we used a finite-difference method on a $1024^2$ grid with $\delta t = 0.005$ and $\delta x = 0.2$ and using $\gamma =0.1$ and $\kappa = 0.5$. For $N=2$ with adiabatic approximations this entails solving a set of two coupled partial differential equations ({\it cf.} eq. \ref{SA}, with $d/dt\rightarrow \partial/\partial t -k\nabla^2$);
for the full $N=2$ there are five (eq. \ref{N2fullmodel}); for $N \rightarrow \infty$ there are seven
(eq. \ref{Ninfeqs}, where we do not evolve the ${\cal P}$ and $\rho$ equations since they have no bearing on the net chirality and do not affect the dynamics of the other $7$ equations);
and for $N=5$ there are seventeen coupled partial differential equations (not shown). Each system was initially prepared in a near racemic phase with a $10^{-5}$ asymmetry in initial monomer concentrations. All other polymer concentrations were initially set to zero.  

Figure \ref{fig:Ncompare} shows the temporal evolution of the spatially-averaged net chirality, $\left\langle \theta (t) \right\rangle$, for $N=2$ (with and without adiabatic approximations), $N =5$, and $N \rightarrow \infty$. One can see that for all values of $N$, $\left\langle \theta (t)\right\rangle$ approaches a constant for large times, indicating that steady-state conditions have been achieved. It is also clear that {\it the asymptotic values of $\left\langle \theta(t)\right\rangle$,  $\theta{ss}$, increase with increasing $N$}: for $N = 2$ (full model), $\theta_{ss} \rightarrow 0.83$; for $N=5$, $\theta_{ss} \rightarrow 0.97$; and for $N \rightarrow \infty$, $\theta_{ss} \rightarrow 1$. 

\begin{figure}
\centerline{\includegraphics[width=4in,height=3in]{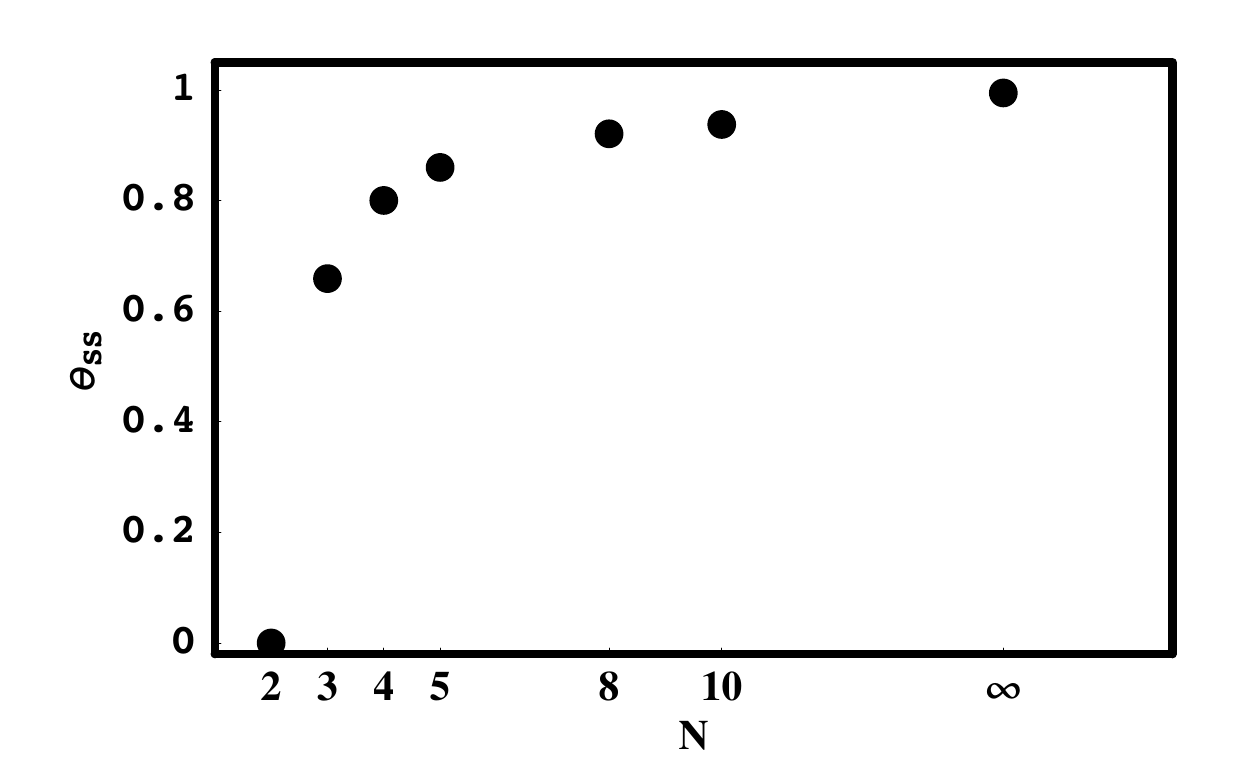}}
\caption{Spatially-averaged, steady-state value of the net chirality, $\theta_{ss}$, for various maximum polymer lengths $N$, with $\kappa=0.5$ and $\gamma=0.3$. Since $\gamma > \gamma_c \mid_{N=2} = 0.25$, $\theta_{ss}$ for $N=2$ is zero. Note that as $N$ increases so does the net chirality, with $\theta \rightarrow 1$ for $N \rightarrow \infty$.} \label{fig:ThetaN}
\end{figure}

\begin{figure}
\centerline{\includegraphics[width=4in,height=3in]{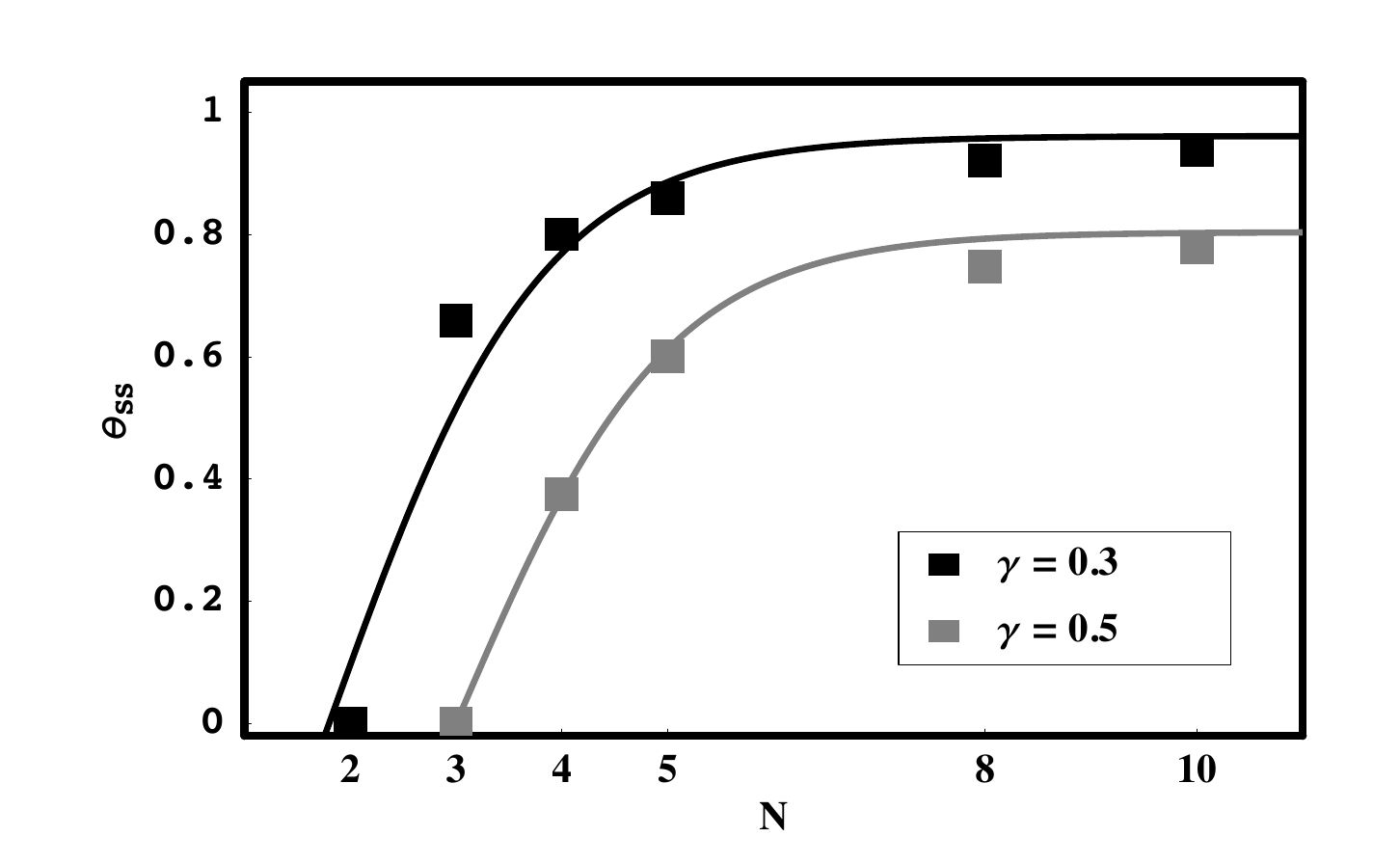}}
\caption{Fit of steady-state chirality, $\theta_{ss}$, for models with various maximum polymer lengths $N$, with $\kappa=0.5$ and $\gamma=0.3$ (black curve and circles) and $\gamma=0.5$ (grey curve and circles).  The fits (continuous curves) are given in eq. \ref{thetafit}.  Since $\gamma > \gamma_c \mid_{N=2} = 0.25$, $\theta_{ss}$ for $N=2$ is zero. The same for $N=3$ when $\gamma=0.5$.} \label{fig:ThetaFit}
\end{figure}

To study the spatiotemporal steady-state solutions in more detail, we examined the behavior of systems initially prepared in a homochiral phase ($\left\langle \theta(t=0)\right\rangle=1$) for different values of  $\gamma$ for models with several values of maximum polymer length $N$, and for $N \rightarrow \infty$. (Again on a $1024^2$ grid with $\delta t = 0.005$ and $\delta x = 0.2$.) That is, we examined whether homochirality is a stable equilibrium solution of the reaction network, and how this stability depends on both the maximum polymer length $N$ and autogenic reaction rate $\gamma$. The results are shown in figures \ref{fig:ThetaN} and \ref{fig:ThetaFit}. From figure \ref{fig:ThetaN}, we can immediately see that homochirality is only strictly stable as $N\rightarrow \infty$: for all other $N$, $\theta_{ss}$ drifts away from unity when perturbed. However, even for models with $N$ as low as $N=4$, $\theta_{ss}\gtrsim 0.8$, or within only $20\%$ of homochirality. 
For $N=2$, $\gamma =0.3$ is larger than the critical value ($\gamma_c \mid _{N=2} = 0.25$, see section \ref{reducedmodel}) so the system is stable as a racemate. We have verified that this trend is true for any $\gamma > 0$ so long as $\gamma < \gamma_c\mid_{N \rightarrow \infty} = \frac{\sqrt{2}}{2}$. Independently of $N$, for $\gamma =0$ all systems achieve homochirality, while for $\gamma > \gamma_c\mid_{N \rightarrow \infty}$ racemates are the only stable solution. (Recall, however, that in the $N\infty$ limit, the homochiral solutions are unstable and unphysical, as discussed in section 4.) Therefore, {\it there exists a range of} $\gamma$, $ 0 < \gamma < \frac{\sqrt{2}}{2} $, {\it where the net chiral excess is dependent on} $N$. Using the nominal values for reaction coefficients introduced in section \ref{N2noAdiabatic}, this bound translates to $0 < \varepsilon < 2.5 \times 10^{-6}{\rm s}^{-1}$. 

Figure \ref{fig:ThetaFit} shows that the dependence of $\theta_{ss}$ on both $N$ and $\gamma$ satisfies the approximate fit,

\begin{eqnarray}
\theta_{ss} = (1 - (\gamma_c/ \gamma)^3)^{1/2} \tanh( aN - b \gamma),
\label{thetafit}
\end{eqnarray}
where $a$ and $b$ are $\gamma$-dependent fitting  constants, and we used $\gamma_c=\gamma_c\mid_{N \rightarrow \infty}=\frac{\sqrt{2}}{2}$. For $\gamma < 0.5$, covering most of the range of interest, $a=0.5$ and $b=3$ give quite accurate results. Figure \ref{fig:ThetaFit} shows the curves for $\gamma =0.3$ and $\gamma=0.5$. 

\begin{figure}
\centerline{\includegraphics[width=5in,height=3in]{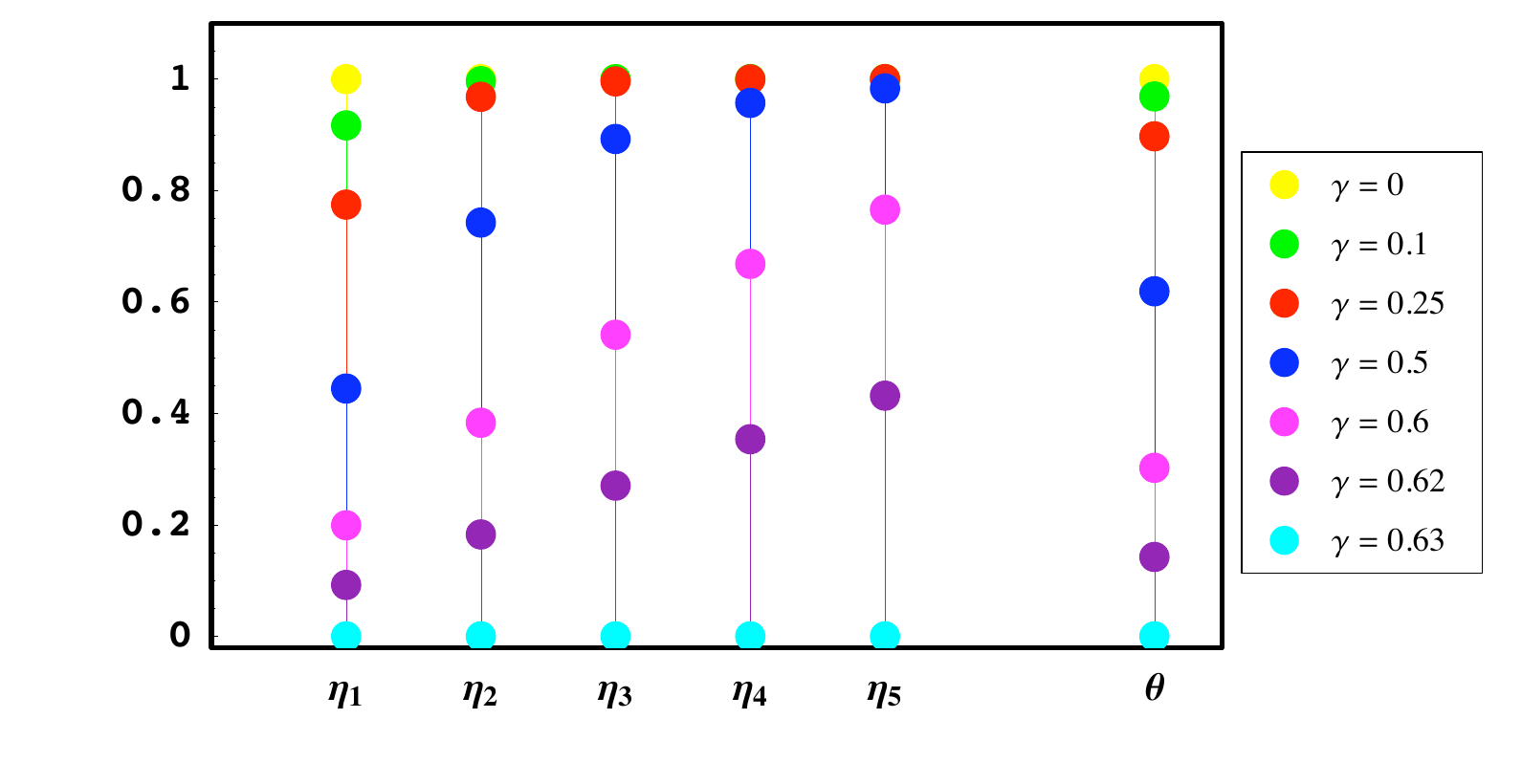}}
\caption{Steady-state values for the chiral asymmetry $\eta_n(t)$ for $N=5$, with $1\leq n\leq 5$. The stationary value of $\theta$ is also shown at the far right. Larger polymers achieve a higher degree of chiral purity. Also, the net chirality per polymer is suppressed by increasing $\gamma$. Note that for $\eta_{n}=1$ some points overlay with others.} \label{fig:N5Stationary}
\end{figure}

A related question is whether the increase in chiral excess with increasing $N$ is dominated by longer or shorter homochiral chains. The answer to this question will shed light on the actual mechanisms that allow for greater chiral purity for large $N$ systems. In the context of their spatially-independent model without autogenic monomer production, WC found that the total chiral purity of all polymer chains is greater than that of monomers. Here, we study in detail how $\eta_n$ changes with increasing $n$ and varying $\gamma$. We obtained the steady state values of the net chirality of individual polymer lengths for the set of ordinary differential equations governing the evolution of the $N=5$ system. The results are shown in the stem diagram of figure \ref{fig:N5Stationary}. One can see that for fixed $\gamma$, the longest polymers achieve the largest chiral excess. On the other hand, it is also clear that homochiral growth is inhibited by large autogenic reaction rates.

Sandar's concluded that as $N$ increases bifurcation happens more readily due to the increased number of opportunities for the minority enantiomer to be removed by cross-inhibition \cite{Sandars03}.  This conclusion was used to explain why large $N$ systems have less stringent requirements on the feedback fidelity in order for bifurcation to occur. Although we have not discussed the effects of varying fidelity here, presumably this can also explain why, with the racemizing pressure created by $\gamma$, it is only in the limit $N \rightarrow \infty$ that $\theta \rightarrow 1$. In order for the net chirality to increase, the rate of chiral amplification through enantiomeric cross-inhibition and autocatalysis must exceed the losses from racemizing pressures. In previous sections, we have described a competition between the racemizing pressure due to $\gamma$ and the pressures favoring chirality due to joint enantiomeric feedback and cross-inhibition. The ensuing dynamics are not only dependent on the relative reaction rates, $\varepsilon$, $k_S$ and $k_C$, but also on the length of the longest polymers formed through $C_L$ and $C_R$. The presence of these terms allows for polymers to accelerate the formation of monomers of the same chirality while inhibiting the formation of polymers of the opposite chirality. Due to the increased opportunities for removal of the minority enantiomer and for enantiomeric autocatalysis provided by large $N$ systems, it is therefore only in the limit of large $N$ that the enantiomeric pressures can overcome the racemizing pressure of a (relatively) large $\gamma$.

\section{Conclusion}\label{end}

We know little of the prebiotic conditions that led to first life in early Earth and even less about other possible life-bearing planetary platforms in this and other stellar systems. The first amino acids and sugars may have been formed here, or may have been fed from outer-space; in either case, there may have been a net initial chirality in their concentrations, or the initial conditions may have been racemic. The substrate (or substrates) may or may not have supported autogenic production of monomers, with or without a chiral bias. Given these uncertainties, it is important to analyze in detail the polymerization dynamics of different reactor pools in order to compute the final net chirality produced. Our impetus for introducing autogenic (non-catalytic) monomer prodution to the model has been to simplify the initial conditions, thereby allowing for a truly bottom-up approach to the origin of prebiotic homochirality. We have found that this mechanism provides for a very rich model of chiral evolution whereby the net chirality is dependent upon the autogenic reaction rate parameters $\gamma_L$ and $\gamma_R$ through the length $N$ of the longest polymers formed. We have shown that the presence of these terms allows for a chiral-symmetry breaking phase transition to occur: for symmetric autogenic production, where $\gamma_L=\gamma_R=\gamma$, we found that, for $\gamma > \gamma_c(N)$, where $\gamma_c(N)$ is an $N$-dependent critical value, racemic solutions are the only equilibrium solutions. In other words, efficient autogenic production of monomers from the substrate creates a racemizing pressure that overwhelms the tendency toward homochirality from autocatalysis with enantiomeric cross-inhibition: if this production is too efficient, the average polymer length is strongly suppressed ({\it cf.} figure \ref{fig:PolymerLengths}), and homochirality cannot be achieved. For $2\leq N< \infty$, we found $0.25\leq \gamma_c \leq \sqrt{2}/2$. Using eq. \ref{gammas}, we can express this bound in terms of the various reaction rates in the model. Adopting a similar procedure, equivalent bounds can be derived for any set of polymerization reactions featuring autogenic production terms.

Our results imply that although in the presence of moderate autogenic monomer production homochirality is achievable starting simply from a substrate, it is preferably achieved for reactor pools allowing the formation of large polymers. We found that this result can be expressed by an approximate relation for the steady-state value of the net chirality, $\theta_{ss}\sim \tanh[N]$. This has implications for studying autocatalytic polymerization networks that display chiral symmetry breaking in the laboratory: if non-catalytic production of monomers is active, only systems which allow for the formation of large $N$ will support bifurcation toward significant chiral purity. In principle at least, it should be possible to test the onset of chiral symmetry breaking in autocatalytic systems such as the Soai reaction by varying external parameters affecting the reaction rates. 

Extrapolating from what we have learned thus far in our solar system and elsewhere, the staggering diversity of potential life-bearing planetary platforms spread throughout the galaxy implies in an equivalently staggering diversity of prebiotic conditions that possibly led to chiral life. Searching for general, model-independent results and trends is thus of the utmost importance.

The authors were partially supported by a National Science Foundation grant PHY-0653341. We had access to the NCSA Teragrid cluster under grant number PHY-070021. We acknowledge extensive use of the FermiQCD parallelization program for solving the coupled systems of PDES.


\begin{thebibliography}{}

\bibitem[\protect\citeauthoryear{Blackmond}{2004}]{Blackmond04}
Blackmond, D. G.:2004,
\newblock {Asymmetric autocatalysis and its implications for the
origin of homochirality}.
\newblock {\em PNAS} {\bf 101}, 5732--5736.

\bibitem[\protect\citeauthoryear{Bonner}{1996}]{Bonner}
Bonner, W.~A.:1996,
\newblock {The Quest for Chirality}.
\newblock In David . D. Cline, editor,
{\em Physical Origin of Homochirality in Life}, Santa Monica, California,
February 1995. AIP Conference Proceedings 379, AIP Press, New York.

\bibitem[\protect\citeauthoryear{Brandenburg {\it et al.}}{2004}]{BAHN}
Brandenburg, A., Andersen, A., Nilsson, M., and H\"ofner, S., T.:2005,
\newblock {Homochiral Growth through Enantiometric Cross-Inhibition}.
\newblock {\em Orig. Life Evol. Biosph}, {\bf 35}, 225--241.

\bibitem[\protect\citeauthoryear{Brandenburg and Multam\"aki}{2004}]{BM}
Brandenburg, A. \& Multam\"aki, T.:2004,
\newblock {How Long Can Left and Right Handed Life Forms Coexist?}.
\newblock {\em Int. J. Astrobiol.}, {\bf 3}, 209--219.

\bibitem[\protect\citeauthoryear{Dunitz}{1996}]{Dunitz}
Dunitz, J. D.:1996,
\newblock {Symmetry Arguments in Chemistry}.
\newblock {\em PNAS} {\bf 93}, 14260--14266.

\bibitem[\protect\citeauthoryear{Engel}{1997}]{Engel97}
Engel, M. H. and Macko, S. A.;1997,
\newblock {Isotopic evidence for extraterrestrial non-racemic
amino acids in the Murchison meteorite}.
\newblock {\em Nature} {\bf 389}, 265--268.

\bibitem[\protect\citeauthoryear{Fitz {\it et al.}}{2007}]{Fitz}
Fitz, D., Reiner, H., Plankensteiner, K., and Rode, B. M.;2007,
\newblock {Possible Origins of Biohomochirality}.
\newblock {\em Curr. Chem. Biol.} {\bf 1}, 41--52.

\bibitem[\protect\citeauthoryear{Frank}{1953}]{Frank53}
Frank, F.~C.:1953,
\newblock {On Spontaneous Asymmetric Catalysis}.
\newblock {\em Biochim. Biophys. Acta}, {\bf 11}, 459--463.

\bibitem[\protect\citeauthoryear{Gleiser and Thorarinson}{2006}]{GT}
Gleiser, M. and Thorarinson, J.:2006,
\newblock {Prebiotic homochiralirty as a critical phenomenon}.
\newblock {in press \em Orig. Life Evol. Biosph.}, {\bf 36}, 501--505.

\bibitem[\protect\citeauthoryear{Gleiser}{2007}]{G}
Gleiser, M.:2007,
\newblock {Asymmetric Spatiotemporal Evolution of Prebiotic Homochirality}.
\newblock {\em Orig. Life Evol. Biosph.}, {\bf 37}, 235--251.

\bibitem[\protect\citeauthoryear{Gleiser {\em et al.}}{2008}]{GTW}
Gleiser, M., Thorarinson, J., and Walker, S.I.:2008,
\newblock {Punctuated Chirality}.
\newblock {\em arXiv:astro-ph/0802.1446. Submitted for publication.}

\bibitem[\protect\citeauthoryear{Goldenfeld}{1992}]{Goldenfeld}
Goldenfeld, N.:1992,
\newblock {\em Lectures on Phase Transitions and the Renormalization Group}.
\newblock Addison Wesley, New York.

\bibitem[\protect\citeauthoryear{Gunton {\em et al.}}{1983}]{Gunton}
Gunton, J. D., San Miguel, M., and Sahni, P. S.:1983,
\newblock {In C. Domb and J. L. Lebowitz, editors {\em Phase Transitions
and Critical Phenomena v. 8},}
\newblock Academic Press, London.

\bibitem[\protect\citeauthoryear{Haken}{1983}]{Haken}
Haken, H.:1983,
\newblock {\em Synergetics: An Introduction}.
\newblock Springer-Verlag, Berlin.

\bibitem[\protect\citeauthoryear{Joyce}{1984}]{Joyce}
Joyce,~G.~F. {\em et al} :1984,
\newblock {Chiral Selection in Poly(C)-directed Synthesis of Oligo(G)}.
\newblock {\em Nature}, {\bf 310}, 602.

\bibitem[\protect\citeauthoryear{Kondepudi and Nelson}{1985}]{KN85}
Kondepudi,~D.~K. and Nelson, G.~W.:1985,
\newblock {Weak Neutral Currents and the Origin of Biomolecular Chirality}.
\newblock {\em Nature}, {\bf 314}, 438--441.

\bibitem[\protect\citeauthoryear{Langer}{1992}]{Langerrev}
Langer, J. S.:1992,
\newblock {An introduction to the kinetics of first-order phase transitions}.
\newblock {In C. Godr\`eche, editor,
{\em Solids Far from Equilibrium}, (Cambridge University Press, Cambridge).}

\bibitem[\protect\citeauthoryear{Nilsson  {\it et al.}}{2004}]{NBAH}
Nilsson, M., Brandenburg, A., Andersen, A., and H\"ofner, S. .:2005,
\newblock {Unidirectional polymerization leading to homochirality in the RNA world}.
\newblock {\em Int. J. Astrobiology}, {\bf 4}, 233--239.

\bibitem[\protect\citeauthoryear{Pizzarello and Cronin}{1998}]{Cronin98}
Pizzarello, S. and Cronin, J. R.:1998,
\newblock {Alanine enantiomers in the Murchison meteorite}.
\newblock {\em Nature} {\bf 394}, 236.

\bibitem[\protect\citeauthoryear{Saito and Hyuga}{2004}]{SH}
Saito, Y. and Hyuga, H.:2004,
\newblock {Chirality Selection Models in a Closed System}.
\newblock {arXiv.org:physics/0408105}.

\bibitem[\protect\citeauthoryear{Sandars}{2003}]{Sandars03}
Sandars, P.~G.~H.:2003,
\newblock {A Toy Model for the Generation of Homochirality
During Polymerization}.
\newblock {\em Orig. Life Evol. Biosph.}, {\bf 33}, 575--587.

\bibitem[\protect\citeauthoryear{Sandars}{2005}]{Sandars05}
Sandars, P.~G.~H.:2005,
\newblock {Chirality in the RNA World and Beyond}.
\newblock {\em Int. J. Astrobiology}, {\bf 4}, 49--61.

\bibitem[\protect\citeauthoryear{Soai}{1995}]{Soai}
Soai, K., Shibata, T., Morioka, H.
and Choji, K.:1995,
\newblock {Asymmetric autocatalysis and amplification of enantiometric
excess of a chiral molecule}.
\newblock {\em Nature} {\bf 378}, 767--768.

\bibitem[\protect\citeauthoryear{Wattis and Coveney}{2005}]{WC}
Wattis, J.~A. and Coveney, P.~V.:2005,
\newblock {Symmetry-Breaking in Chiral Polymerization}.
\newblock {\em Orig. Life Evol. Biosph.}, {\bf 35}, 243--273.

\bibitem[\protect\citeauthoryear{Yamagata}{1966}]{WNC0}
Yamagata, Y.:1966,
\newblock {A hypothesis for the asymmetric appearance of biomolecules on earth}
\newblock {\em J. Theoret. Biol.} {\bf 11}, 495--498.


\end{thebibliography}
\end{document}